\documentclass[journal=jctcce,manuscript=article]{achemso}

\usepackage{amsmath}
\usepackage{amssymb}
\usepackage{amsfonts}
\usepackage{bm}
\usepackage{braket}
\usepackage{float}
\usepackage{graphicx}
\usepackage{mathtools}
\usepackage{listings}
\usepackage[Symbol]{upgreek}
\usepackage{subcaption}
\usepackage{changes}

\DeclareMathOperator{\el}{\mathrm{e}^{--}}
\DeclareMathOperator{\pr}{\mathrm{p}^{+}}

\title{Quantum Proton Effects from Density Matrix Renormalization Group Calculations}

\author{Robin Feldmann}
\affiliation{ETH Z\"urich, Laboratory of Physical Chemistry, Vladimir-Prelog-Weg 2, 8093 Z\"urich, Switzerland}

\author{Andrea Muolo}
\affiliation{Fritz Haber Center for Molecular Dynamics, Institute of Chemistry, The Hebrew University of Jerusalem, Jerusalem 91904, Israel}

\author{Alberto Baiardi}
\affiliation{ETH Z\"urich, Laboratory of Physical Chemistry, Vladimir-Prelog-Weg 2, 8093 Z\"urich, Switzerland}

\author{Markus Reiher}
\email{markus.reiher@phys.chem.ethz.ch}
\affiliation{ETH Z\"urich, Laboratory of Physical Chemistry, Vladimir-Prelog-Weg 2, 8093 Z\"urich, Switzerland}

\begin{document}

\begin{abstract}

We recently introduced [J. Chem. Phys. 152 2020, 204103] the nuclear-electronic all-particle density matrix renormalization group (NEAP-DMRG) method to solve the molecular Schr\"{o}dinger equation, based on a stochastically optimized orbital basis, without invoking the Born-Oppenheimer approximation.
In this work, we combine the DMRG method with the nuclear-electronic Hartree-Fock (NEHF-DMRG) approach, treating nuclei and electrons on the same footing.
Inter- and intra-species correlations are described within the DMRG method without truncating the excitation degree of the full configuration interaction wave function.
We extend the concept of orbital entanglement and mutual information to nuclear-electronic wave functions and demonstrate that they are reliable metrics to detect strong correlation effects.
We apply the NEHF-DMRG method to the HeHHe$^+$ molecular ion, to obtain accurate proton densities, ground-state total energies, and vibrational transition frequencies by comparison with state-of-the-art data obtained with grid-based approaches and modern configuration interaction methods.
For HCN, we improve on the accuracy of the latter approaches with respect to both the ground-state absolute energy and proton density which is a major challenge for multi-reference nuclear-electronic state-of-the-art methods.

\end{abstract}

\maketitle

\section{Introduction}
\label{sec:intro}

Although methods of many-particle quantum mechanics relying on the omnipresent Born--Oppenheimer (BO) approximation are cornerstones of quantum chemistry and molecular physics, the recent advances of methods that do not rely on the BO approximation highlight the relevance of nuclear quantum and non-adiabatic effects in many chemical phenomena, including hydrogen tunneling, vibrationally excited states, and proton transfer reactions \cite{Manby2020_EqualFooting,Manby2020_Coupling,Vendrell2020_PesFree,Muolo2020_Nuclear,Brorsen2020_multicomponentCASSCF,Brorsen2020_SelectedCI-PreBO,Adamowicz2020_AtomMol,Hammes-Schiffer2020_Review,Gross_2020GreensFunc,Rubio2021_woBO}.

In nuclear-electronic methods, nuclei are treated on an equal footing with electrons. 
As a consequence, approximations of eigenfunctions of the nuclear-electronic Hamiltonian include nonadiabatic couplings between nearly degenerate electronic states systematically
\cite{BornHuang1954,Wolniewicz1964_accurate,Gauss2006_dboc,Komasa_2009nonadiabatic},
and the zero-point vibrational energy level is obtained in a single calculation.
This avoids the need to approximate different potential energy hypersurfaces, which is a prohibitive task already for systems comprising more than a few atoms \cite{Behler2007_PES-NN,Truhlar2013_PES-N4,Dawes2012_GlobalPES-Ozone}.
A salient feature of the nuclear-electronic Hamiltonian is that it only contains one- and two-body terms as opposed to the BO vibrational Hamiltonian that contains complicated couplings of arbitrary higher order.\cite{Christiansen2004_SecondQuantization}
The foremost challenge of nuclear-electronic methods is to include the nuclear-electronic correlations reliably and to effectively identify and screen the orbital active space for all particle types. 

The nuclear-electronic wave function ansatz consists of a linear combination of either explicitly correlated basis functions \cite{Adamowicz2012_Review} or (anti)symmetrized products of single-particle functions \cite{Nakai1998_NOMO-Original,Hammes-Schiffer2020_Review}, that is, the orbitals.
Approaches based on the former have been successfully applied to calculate highly accurate energies of small molecules, but are limited by the factorial scaling of the (anti)symmetrization operators \cite{Adamowicz2012_Review,Adamowicz2013_rev} that prevents systems larger than eight-particle systems, such as BH, from being studied \cite{Adamowicz2020_AtomMol}.
Conversely, this problem is absent for (anti)symmetric products of orbitals, for which
the Slater--Condon or L{\"o}wdin rules can be exploited so that the corresponding methods are applicable to much larger systems.
The first method based on nuclear and electronic orbitals was introduced in a series of articles by Thomas in 1969 \cite{Thomas1969_protonic1,Thomas1969_protonic2,Thomas1970_protonic3,Thomas1971_protonic4}. Pettitt developed \cite{Pettitt1986_FirstMcHF} in 1986 the first nuclear-electronic Hartree--Fock (NEHF) with self-consistent field (SCF) optimization, which in 1987 was employed to compute proton densities\cite{Pettitt1987_ScfProton}.
Nakai et al.\ in 1998, were the first to develop a practical theory within the molecular-orbital theory, called the NOMO method \cite{Nakai1998_NOMO-Original}.
Since then, various orbital-based SCF approaches have been developed under different names, that is, NEO \cite{Hammes-Schiffer2002}, ENMO \cite{Valeev2004_ENMO}, MCMO \cite{Tachikawa2002}, and APMO \cite{APMO_2008}.
For inter- and intra-species correlation effects, post-HF methods have been generalized to nuclear-electronic wave functions \cite{HammesSchiffer2004_Tunneling-Correlation}, 
such as the second-order M{\o}ller--Plesset perturbation theory \cite{nakai2007nuclear}, coupled cluster singles and doubles (CCSD) \cite{Nakai2003,Ellis2016_Development,Pavosevic2019_MulticomponentCC,Pavosevic2021_MulticomponentUCC,Hammes-Schiffer2021_MulticomponenDF}, and configuration interaction (CI) methods \cite{Valeev2004_ENMO,Hammes-Schiffer2002,Patrick_2015_EN-MFCI,Brorsen2020_SelectedCI-PreBO,Brorsen2020_multicomponentCASSCF}.
In parallel, also the development of nuclear-electronic density functional theory is also progressing \cite{Hammes-Schiffer2012_MulticomponentDFT,Hammes-Schiffer2017_MulticomponentDFT,Brorsen2018_TransferableDFT,HammesSchiffer2019_GGA-DFT}.

While nuclear-electronic methods have been primarily based on the single-reference ansatz, the development of multi-reference methods has been hindered by the steep factorial scaling of the number of many-particle basis functions in the full orbital space or a subset of it [as in complete active space (CAS) approaches] \cite{Hammes-Schiffer2002,Hammes-Schiffer2005_NOCI}.
An example of a multi-reference method is the nuclear-electronic variant of the heat-bath configuration interaction (HBCI) method \cite{Holmes2016_HBCI,Sharma2017_HBCI} developed by Fajen and Brorsen \cite{Brorsen2020_SelectedCI-PreBO,Brorsen2020_multicomponentCASSCF}.
The nuclear electronic HBCI method has been developed as a CI method with single, double, triple, and quadruple excitations (CISDTQ) and CAS-SCF approach which has been applied to systems with up to 21 particles.
In our previous work \cite{Muolo2020_Nuclear}, we address the factorial scaling problem in the full orbital space with the density matrix renormalization group (DMRG) optimization, extended to systems composed of different types of quantum particles \cite{Muolo2020_Nuclear}.
The DMRG method scales polynomially with respect to the number of orbitals owing to the matrix product state (MPS) factorization of the molecular wave function \cite{White1992,white1993density,Chan2008_Review,Zgid2009_Review,Marti2010_Review-DMRG,schollwock2011density,chan2011density,Wouters2014_Review,Kurashige2014_Review,Olivares2015_DMRGInPractice,szalay2015tensor,Yanai2015,Baiardi2020_Review}.
We employed\cite{Muolo2020_Nuclear} a stochastically sampled all-particle wave function ansatz composed by variationally optimized non-orthogonal nuclear-electronic molecular orbitals. 
The resulting optimization algorithm was called the nuclear-electronic all-particle (NEAP) method.
The acronym 'NEAP' explicitly emphasizes
that all nuclei are considered as quantum nuclei, whereas we will denote the case of {\it selected nuclei} to be treated
quantum mechanically by the acronym 'NE' which denotes 'nuclear-electronic'.

The NEAP algorithm relies on a fully stochastic optimization of all wave function parameters, a task that becomes challenging for large basis sets.
This work provides a computationally less demanding method by combining DMRG with molecular orbitals that are expressed as a linear combination of pre-optimized basis functions and the molecular orbital coefficients are optimized with NEHF.
We consider the case of spin-restricted electrons and spin-unrestricted nuclei for molecules with an electronic closed-shell ground-state configuration.
The nuclei are well separated in space and highly localized in comparison with electrons and, hence, the unrestricted ansatz is a suitable one \cite{Valeev2004_ENMO}.

By comparison with data obtained for HeHHe$^+$ and HCN with the nuclear-electronic HBCI(SDTQ) method by Brorsen \cite{Brorsen2020_SelectedCI-PreBO}, we show that NEHF-DMRG can efficiently describe the electron-proton correlations in the ground state and yield accurate proton densities.
We elaborate the concept of orbital entanglement and mutual information for the nuclear-electronic case and we study how they vary for canonical HF orbitals and natural orbitals (NOs). 
We note here that Schilling et al.\ recently pointed out that the actual quantity measured by the von Neumann entropy is not solely the entanglement, but instead a mixture of classical correlation and entanglement, termed total correlation \cite{Schilling2020_entanglement,Schilling2020_correlation}.
However, since the term correlation has a different meaning in quantum chemistry and for the sake of consistency with the previous literature, 
\cite{Legeza2003_OrderingOptimization,White_2006information,Boguslawski2012_entanglement,Stein2016_AutomatedSelection}
we will refer to the total correlation as orbital entanglement.

This work is organized as follows.
Sec.~\ref{sec:theory} introduces the Hamiltonian and wave function ansatz.
Sec.~\ref{sec:mc_hf_method} presents the working equations of the nuclear-electronic Hartree--Fock method and Sec.~\ref{sec:2nd-quant} the Hamiltonian and the FCI wave function in second quantization.
Afterwards, we review the DMRG algorithm in Sec.~\ref{sec:mc-dmrg}.
In Sec.~\ref{sec:mutinf}, we define orbital entanglement for nuclear-electronic wave functions.
The computational details and the results obtained for HeHHe$^+$ and HCN are presented in Sec.~\ref{sec:results}.

\section{Theory}
\subsection{Nuclear-Electronic Hamiltonian and Wave function}
\label{sec:theory}

Consider a system composed of $N_\mathrm{t}$ different types of particles in the presence of $N_\text{c}$ external point charges $Q_b$, with $b=1,\dots,N_\text{c}$. 
$N_i$ is the number of particles of type $i$ with the corresponding masses $m_i$ and charges $q_i$. 
$\mathbf{r}_{ia},\mathbf{R}_b\in\mathbb{R}^3$ are position vectors for the $a$-th particle of type $i$ and point charge $b$, respectively. 
The non-relativistic nuclear-electronic Hamiltonian for such a system in Hartree atomic units and in the position representation reads 
\begin{equation}
\begin{aligned}
 \mathcal{H} = \sum_{i}^{N_t} \sum_{a}^{N_i} h(\mathbf{r}_{i,a}) 
 +\frac{1}{2} \sum_{ij}^{N_t} \sum_{a}^{N_i} 
 \sum_{ \substack{b \\ \text{if} \, i=j \\ \text{then} \,a\ne b} }^{N_j} g(\mathbf{r}_{i,a},\mathbf{r}_{i,b})~,
\end{aligned}
\label{eq:PreBO-Hamiltonian}
\end{equation}
with one-body and two-body operators defined as
\begin{align} 
 & h(\mathbf{r}_{i,a})  = - \frac{1}{2 m_i} \bm{\nabla}^2_{i,a}
 + \sum_{b}^{N_{\text{c}}}\frac{q_i Q_b}{|\mathbf{r}_{i,a} -\mathbf{R}_b|} ~,
 \label{eq:1body_H}
\end{align} 
and 
\begin{align} 
 & g(\mathbf{r}_{i,a},\mathbf{r}_{j,b}) = \frac{q_i q_j}{|\mathbf{r}_{i,a} -\mathbf{r}_{j,b}|} ~,
 \label{eq:2body_H}
\end{align} 
respectively. $\bm{\nabla}^2_{i,a}$ is the Laplace operator acting on the $a$-th particle of type $i$.
The one-body operator, $h$, in Eq.~(\ref{eq:1body_H}) consists of the kinetic-energy operator and the Coulomb interaction of quantum particles and classical point charges (e.g., atomic nuclei heavier than the proton).
The two-body operator, $g$, in Eq.~(\ref{eq:2body_H}), describes the Coulomb interaction between quantum particles.

We approximate the ground state of the Hamiltonian in Eq.~(\ref{eq:PreBO-Hamiltonian}) with the HF wave function, $\Psi$.
In contrast to electronic structure theory, $\Psi$ explicitly depends on nuclear and electronic coordinates and is given by
\begin{equation}
 \Psi (\mathbf{r}_{1},\dots,\mathbf{r}_{N_\text{t}}) 
 = \prod_i^{N_{\text{t}}} \Phi_{i} (\mathbf{r}_{i}) ~,
 \label{eq:ansatz}
\end{equation}
where $\Phi_{i}$ is a Slater determinant for all particles of type $i$ and $\mathbf{r}_i$ collects their position vectors.
Each of these Slater determinants for particle type $i$ is constructed from a set of spatial molecular orbitals, $\varphi_{is, \mu}$, that is expressed as linear combinations of $L_i$ (nuclear or electronic) Gaussian orbitals (LCGOs)
\begin{equation}
 \varphi_{is, \mu} (\mathbf{r}_{i,a}) = \sum_p^{L_i} c_{i s, \mu p} ~ \chi_{i,p}(\mathbf{r}_{i,a}) ~.
 \label{eq:LCAO}
\end{equation}
where $\chi_{i,p}(\mathbf{r}_{i,a})$ are (contracted) Gaussian type orbitals [(c)GTOs], defined by their Gaussian width, shift, angular momentum and possibly pre-optimized contraction coefficients.
The corresponding spin orbital, $\phi_{is, \mu}$, is obtained by multiplying the spatial orbital, $\varphi_{is, \mu}$, with an appropriate spin function.
The molecular and Gaussian orbitals are written in the Dirac notation as $\varphi_{is, \mu} (\mathbf{r}) \rightarrow \ket{\mu_{is}}$ and $\chi_{i,p} (\mathbf{r}) \rightarrow \ket{p_{i}}$, respectively.

In the following, if not stated otherwise, $i,j$ correspond to particle types, running from 1 to $N_\mathrm{t}$, lower case Greek indices $\mu,\nu$ refer to molecular orbitals, $s=\uparrow,\downarrow$ to the fermionic spin projection, and $p,q,r,t$ label different cGTOs.

\subsection{Nuclear-Electronic Hartree--Fock Method}
\label{sec:mc_hf_method}

\subsubsection{Unrestricted Nuclear-Electronic Hartree--Fock Equations}

The nuclear-electronic Hartree--Fock equations follow from the minimization of the molecular spin-orbital Lagrangian functional
\begin{equation}
 \mathcal{L}[\{\phi_{is, \mu}\}] = E[\{\phi_{is, \mu}\}] - \sum_i^{N_\text{t}} \sum_{\mu\nu}^{N_i}
 \sum_{ss^\prime=\uparrow,\downarrow} \epsilon_{iss^\prime,\mu\nu}\left(\Braket{\mu_{is}|\nu_{is^\prime}} - \delta_{\mu\nu} \delta_{ss^\prime}\right)~,
 \label{eq:lagrangian_hf_main}
\end{equation}
where $\epsilon_{iss^\prime,\mu\nu}$ are Lagrangian multipliers for all orthonormalization constraints.
By writing one-body integrals as $h_{is, \mu}$ and two-body integrals in physics notation as
$\Braket{\mu_{is} \nu_{js^\prime}|\mu_{is} \nu_{js^\prime}}$ and integrating out the spin, we evaluate the energy functional based on the Slater--Condon rules as
\begin{equation}
\begin{aligned}
 E = &\sum_i^{N_\text{t}} \sum_\mu^{N_{i\uparrow}} h_{i\uparrow,\mu}
 + \sum_i^{N_\text{t}} \sum_\mu^{N_{i\downarrow}} h_{i\downarrow,\mu}
 \\
 &+ \frac{1}{2}\sum_i^{N_\text{t}} \sum_{\mu\ne\nu}^{N_{i\uparrow}}
 \Braket{\mu_{i\uparrow}\nu_{i\uparrow}|\mu_{i\uparrow}\nu_{i\uparrow}} -
 \Braket{\mu_{i\uparrow}\nu_{i\uparrow}|\nu_{i\uparrow}\mu_{i\uparrow}} 
 \\
 &+ \frac{1}{2}\sum_i^{N_\text{t}} \sum_{\mu\ne\nu}^{N_{i\downarrow}}
 \Braket{\mu_{i\downarrow}\nu_{i\downarrow}|\mu_{i\downarrow}\nu_{i\downarrow}} -
 \Braket{\mu_{i\downarrow}\nu_{i\downarrow}|\nu_{i\downarrow}\mu_{i\downarrow}} 
 \\ 
 &+ \sum_i^{N_\text{t}} \sum_{\mu}^{N_{i\uparrow}} \sum_{\nu,\,\nu\ne\mu}^{N_{i\downarrow}} 
 \Braket{\mu_{i\uparrow}\nu_{i\downarrow}|\mu_{i\uparrow}\nu_{i\downarrow}} 
 + \sum_{i\neq j}^{N_\text{t}} \sum_{\mu}^{N_{i\uparrow}} \sum_{\nu}^{N_{j\downarrow}}
 \Braket{\mu_{i\uparrow} \nu_{j\downarrow}|\mu_{i\uparrow} \nu_{j\downarrow}}
 \\
 &+ \frac{1}{2} \sum_{i\neq j}^{N_\text{t}} \sum_{\mu}^{N_{i\uparrow}} \sum_{\nu}^{N_{j\uparrow}}
 \Braket{\mu_{i\uparrow} \nu_{j\uparrow}|\mu_{i\uparrow} \nu_{j\uparrow}}
 + \frac{1}{2} \sum_{i\neq j}^{N_\text{t}} \sum_{\mu}^{N_{i\downarrow}} \sum_{\nu}^{N_{j\downarrow}}
 \Braket{\mu_{i\downarrow} \nu_{j\downarrow}|\mu_{i\downarrow} \nu_{j\downarrow}} ~.
 \end{aligned}
 \label{eq:EnergyFunctional}
\end{equation}
In the following, we abbreviate the functional derivative according to
\begin{equation*}
  \frac{\updelta}{\updelta\bra{\kappa_{i\uparrow}}}:\quad\bra{\kappa_{i\uparrow}} \rightarrow \bra{\updelta\kappa_{i\uparrow}}~.
\end{equation*}
By setting the variation of $\mathcal{L}$ to zero and obtaining the variation of the energy functional from Eq.~(\ref{eq:EnergyFunctional}), we write
\begin{equation}
\begin{aligned}
    \frac{\updelta \mathcal{L}}{\updelta\bra{\kappa_{i\uparrow}}}
        =  &~\Braket{\updelta\kappa_{i\uparrow}|h|\kappa_{i\uparrow}}
        + \sum_{\nu}^{N_{i\uparrow}} \Braket{\updelta\kappa_{i\uparrow}\nu_{i\uparrow}|\kappa_{i\uparrow}\nu_{i\uparrow}} - \Braket{\updelta\kappa_{i\uparrow}\nu_{i\uparrow}|\nu_{i\uparrow}\kappa_{i\uparrow}} \\
        &~+ \sum_{\nu}^{N_{i\downarrow}}  \Braket{\updelta\kappa_{i\uparrow}\nu_{i\downarrow}|\kappa_{i\uparrow}\nu_{i\downarrow}}
        + \sum_{j\neq n}^{N_\text{t}} \sum_{\nu}^{N_{j\uparrow}}
        \Braket{\updelta\kappa_{i\uparrow} \nu_{j\uparrow}|\kappa_{i\uparrow} \nu_{j\uparrow}}
        \\ 
        &~+ \sum_{j\neq n}^{N_\text{t}} \sum_{\nu}^{N_{j\downarrow}}
        \Braket{\updelta\kappa_{i\uparrow} \nu_{j\downarrow}|\kappa_{i\uparrow} \nu_{j\downarrow}} 
        - \sum_{\nu}^{N_{i\uparrow}} \epsilon_{i\uparrow,\kappa\nu} \Braket{\updelta\kappa_{i\uparrow}|\nu_{i\uparrow}} \\= &~0~.
\end{aligned}
\label{eq:LagrangeHF}
\end{equation}
In order to find an expression for the Fock operators, we define the Coulomb operator, $\mathcal{J}_{i s, \mu}(1)$, and the exchange operator, $\mathcal{K}_{i s, \mu}(1)$, in analogy to the Hartree--Fock method of electronic structure theory. 
With the two-body interaction, $g$, the Coulomb operator is written as
\begin{equation}
    \mathcal{J}_{j s, \nu}(1)  \ket{\mu_{i\uparrow} (1)} = \left( \int\mathrm{d}^3 r_{j2}\ \left| \varphi_{js,\nu} (2)  \right|^2 g(\mathbf{r}_{i1},\mathbf{r}_{j2}) \right) \ket{\mu_{i\uparrow} (1)} \, ,
    \label{eq:Coulomb}
\end{equation}
and couples spatial orbitals, $\varphi_{is, \mu}$, of different particle types and spins, and the exchange operator reads
\begin{equation}
    \mathcal{K}_{i \uparrow,\nu}(1) \ket{\mu_{i\uparrow} (1)} = \left( \int\mathrm{d}^3 r_{i2}\ \varphi_{i\uparrow,\nu} (2) g(\mathbf{r}_{i1},\mathbf{r}_{i2}) \varphi_{i\uparrow,\mu} (2)\right) \ket{\nu_{i\uparrow} (1)}~,
    \label{eq:Exchange}
\end{equation}
which couples only spatial orbitals of particles of the same type and spin. 
The corresponding matrix elements of the operators read
\begin{equation}
  \Braket{\mu_{i\uparrow}(1)| \mathcal{J}_{j s,\nu}(1) | \mu_{i\uparrow}(1)}
    = \Braket{\mu_{i\uparrow} \nu_{js}|\mu_{i\uparrow} \nu_{js}}~,
  \label{eq:MatrixElementsHF}
\end{equation}
and
\begin{equation}
    \Braket{\mu_{i\uparrow}(1)| \mathcal{K}_{i \uparrow,\nu}(1) | \mu_{i\uparrow}(1)} = \Braket{\mu_{i\uparrow}\nu_{i\uparrow}|\nu_{i\uparrow}\mu_{i\uparrow}}~.
\end{equation}
Based on the definitions of Coulomb, Eq.\ (\ref{eq:Coulomb}), and exchange, Eq.\ (\ref{eq:Exchange}), operators, the variation of the Lagrangian functional, Eq.\ (\ref{eq:LagrangeHF}), reads
\begin{equation}
\begin{aligned}
    \frac{\updelta \mathcal{L}}{\updelta\bra{\kappa_{i\uparrow}}}
        =  &~\bra{ \updelta \kappa_{i\uparrow}(1)} \Bigg{(} h
        + \sum_\nu^{N_{i\uparrow}} \left( \mathcal{J}_{i \uparrow,\nu}(1) -\mathcal{K}_{i \uparrow,\nu}(1)\right) \\
        &~+ \sum_\nu^{N_{i \downarrow}}  \mathcal{J}_{i \downarrow,\nu}(1)
        + \sum_{j\neq n}^{N_\text{t}} \sum_\nu^{N_{j\uparrow}} \mathcal{J}_{j \uparrow,\nu}(1)
        + \sum_{j\neq n}^{N_\text{t}} \sum_\nu^{N_{j \downarrow}} \mathcal{J}_{j \downarrow,\nu}(1) \Bigg{)} \ket{\kappa_{i\uparrow}(1)} \\
        &- \sum_{\nu}^{N_{i\uparrow}} \epsilon_{i\uparrow,\kappa\nu} \Braket{\updelta\kappa_{i\uparrow}|\nu_{i\uparrow}} \\= &~0~.
\end{aligned}
    \label{eq:LagrangeHF2}
\end{equation}
We define in close analogy to electronic structure theory, the Fock operator acting on particles of type $i$ with spin-up as
\begin{equation}
\begin{aligned}
    f_{i\uparrow} =   &~h
        + \sum_\nu^{N_{i\uparrow}} \left( \mathcal{J}_{i \uparrow,\nu}(1) -\mathcal{K}_{i\uparrow,\nu}(1)\right) \\
        &+ \sum_\nu^{N_{i\downarrow}}  \mathcal{J}_{i\downarrow,\nu}(1)
        + \sum_{j\neq i}^{N_\text{t}} \sum_\nu^{N_{j\uparrow}} \mathcal{J}_{j\uparrow,\nu}(1)
        + \sum_{j\neq i}^{N_\text{t}} \sum_\nu^{N_{j\downarrow}} \mathcal{J}_{j\downarrow,\nu}(1)~.
\end{aligned}
        \label{eq:fockop_uhf}
\end{equation}
The Fock operator for spin-down follows by analogy.
A Slater determinant is invariant upon unitary rotations between orbitals of the same particle type, and therefore, the HF equations can be written in a canonical form as 
\begin{equation}
\begin{aligned}
    f_{i\uparrow} \ket{\mu_{i\uparrow}} &= \epsilon_{i\uparrow,\mu} \ket{\mu_{i\uparrow}}~, \\
    f_{i\downarrow} \ket{\mu_{i\downarrow}} &= \epsilon_{i\downarrow,\mu} \ket{\mu_{i\downarrow}}~.
    \label{eq:HFequation}
\end{aligned}
\end{equation}

\subsubsection{Unrestricted Pople--Nesbet-like Equations}

By recalling Eq.~(\ref{eq:LCAO}), we express the spatial part of a molecular spin-orbital as a linear combination of GTOs

\begin{equation}
 \ket{\mu_{is}}    = \sum_p^{L_i} c_{is, \mu p}
     ~\ket{p_i} \, .
  \label{eq:GTO}
\end{equation}
Substituting Eq.~(\ref{eq:GTO}) in the Hartree--Fock equation, Eq.~(\ref{eq:HFequation}), yields
\begin{equation}
  f_{i\uparrow} \ket{\mu_{i\uparrow}}  
    = f_{i\uparrow}   \sum_p^{L_i} c_{i\uparrow,\mu q} \ket{q_i} = \sum_p^{L_i} c_{i\uparrow,\mu q}\, f_{i\uparrow}\ket{q_i}~.
\end{equation}
Next, we multiply by $\bra{p_i}$ from the left to derive the unrestricted nuclear-electronic Fock matrix projected into the GTO basis
\begin{equation}
\begin{aligned}
 F_{i\uparrow,pq} = \bra{p_i} f_{i\uparrow} \ket{q_i} =  &~h_{i,pq} + \sum_\nu^{N_{i\uparrow}} \left( \bra{p_i}\mathcal{J}_{i\uparrow,\nu}\ket{q_i}  -\bra{p_i}\mathcal{K}_{i\uparrow,\nu}\ket{q_i} \right)\nonumber\\
 &~+ \sum_\nu^{N_{i\downarrow}}  \bra{p_i}\mathcal{J}_{i\downarrow,\nu}\ket{q_i}  \nonumber\\
        &~+ \sum_{j\neq i}^{N_\text{t}} \left( \sum_\nu^{N_{j\uparrow}} \bra{p_i}\mathcal{J}_{j\uparrow,\nu}\ket{q_i}
        +\sum_\nu^{N_{j\downarrow}} \bra{p_i}\mathcal{J}_{j\downarrow,\nu}\ket{q_i} \right)~.
\end{aligned}
\end{equation}
We introduce the unrestricted Hartree--Fock density matrix, $\bm{D}_{i\uparrow}$, as
\begin{equation}
  D_{i\uparrow,pq} = \sum_{\mu}^{N_{i\uparrow}} c_{is, \mu p} \, c_{is, \mu q}~,
  \label{eq:Unrestricted_DM}
\end{equation}
write the Coulomb operator expressed in the GTO basis as
\begin{equation}
   \sum_\nu^{N_{j\uparrow}} \bra{p_i} \mathcal{J}_{j\uparrow,\nu} \ket{q_i} =\sum_{rt}^{L_j} D_{j\uparrow,rt} \Braket{p_i r_j|q_i t_j}~,
   \label{eq:Coulomb_FiniteBasis}
\end{equation}
and the exchange operator accordingly
\begin{equation}
    \sum_\nu^{N_{i\uparrow}} \bra{p_i} \mathcal{K}_{i\uparrow,\nu} \ket{q_i} =\sum_{rt}^{L_i} D_{i\uparrow,rt} \Braket{p_i r_i|t_i q_i}~.
    \label{eq:Exchange_FiniteBasis}
\end{equation}
The Fock matrix can then be expressed in terms of density matrices and integrals over GTOs
\begin{equation}
\begin{aligned}
    F_{i\uparrow,pq} = &~h_{i,pq} + \sum_{rt}^{L_i} \left( D_{i\uparrow,rt} \left( \Braket{p_i r_i|q_i t_i} -  \Braket{p_i r_i|t_i q_i} \right) 
    + D_{i\downarrow,rt} \Braket{p_i r_i|q_i t_i} \right) \\
    &~+  \sum_{j\neq i}^{N_\text{t}} \sum_{rt}^{L_j} \left(  D_{j\uparrow,rt} 
        + D_{j\downarrow,rt}  \right) \Braket{p_i r_j|q_i t_j}~.
\end{aligned}
\end{equation}
The unrestricted nuclear-electronic Hartree--Fock equations in the basis of the GTOs read
\begin{equation}
    \sum_q^{L_i} F_{i\uparrow,pq} c_{i\uparrow,\mu q} = \epsilon_{i\uparrow,\mu} \sum_q^{L_i} S_{i,pq} c_{i\uparrow,\mu q}\, ,
    \label{eq:URH}
\end{equation}
where the overlap matrix $\bm{S}_{i}$ is defined as
\begin{equation}
    S_{i,pq} = \Braket{p_i|q_i}.
\end{equation}
Considering all particle types, the resulting Eq.\ (\ref{eq:URH}) forms a set of coupled Pople--Nesbet equations, that is, the unrestricted analogues to the Roothaan--Hall equations \cite{pople1954_uhf}. Accordingly and following Sherril et al. \cite{Valeev2004_ENMO}, we will refer to them as Pople--Nesbet-like equations. In matrix notation they consist of a set of matrix-eigenvalue equations which are coupled via density matrices:
\begin{equation}
    \bm{F}_{is}(\{\bm{D}_{js^\prime}\}) \bm{C}_{is} = \bm{S}_i  \bm{C}_{is}  \bm{E}_{is}\quad \forall ~ij,\,ss^\prime=\uparrow,\downarrow.
\end{equation}
The energy of the corresponding determinant is
\begin{align}
     E = &~\sum_i^{N_\text{t}} \sum_{pq}^{L_i} \left(D_{i\uparrow,pq} 
        + D_{i\downarrow,pq}  \right) h_{i,pq} + \frac{1}{2} \sum_i^{N_\text{t}}  \sum_{pq}^{L_i} \left(D_{i\uparrow,pq} G_{i\uparrow,pq}
        + D_{i\downarrow,pq} G_{i\downarrow,pq} \right),
\end{align}
where $G_{i\uparrow,pq} = F_{i\uparrow,pq} -h_{i,pq}$. 

\subsubsection{Restricted-Unrestricted Nuclear-Electronic Hartree--Fock Method}

In the restricted case, the coefficients for spin-up and spin-down will be set equal, according to
\begin{equation}
    \ket{\mu_{\el\uparrow}} =  \ket{\mu_{\el\downarrow}}.
 \label{eq:restricted}
\end{equation}
With Eq.~(\ref{eq:restricted}), the energy functional defined in Eq.~(\ref{eq:EnergyFunctional}) reads
\begingroup
\allowdisplaybreaks
\begin{align}
    E 
        = &~2\sum_\mu^{N_{\el}/2} h_{\el \mu}
        +
    \sum_{\mu\nu}^{N_{\el}/2}  
    2\Braket{\mu_{\el}\nu_{\el}|\mu_{\el}\nu_{\el}} 
    - \Braket{\mu_{\el}\nu_{\el}|\nu_{\el}\mu_{\el}} \nonumber\\
     &~+      2\sum_{i\neq \el}^{N_\text{t}} \sum_\mu^{N_{\el}/2}
            \sum_{\nu}^{N_{i\uparrow}}
            \Braket{\mu_{\el} \nu_{i\uparrow}|\mu_{\el} \nu_{i\uparrow}}
     +      2\sum_{i\neq \el}^{N_\text{t}} \sum_\mu^{N_{\el}/2}
            \sum_{\nu}^{N_{i\downarrow}}
            \Braket{\mu_{\el} \nu_{i\downarrow}|\mu_{\el} \nu_{i\downarrow}} \nonumber\\
    &~+  \sum_{i\neq\el}^{N_\text{t}} \sum_\mu^{N_{i\uparrow}}
    h_{i\uparrow,\mu}
    + \sum_i^{N_\text{t}} \sum_\mu^{N_{i\downarrow}} 
    h_{i\downarrow,\mu} \nonumber\\
        &~+ \frac{1}{2}\sum_{i\neq\el}^{N_\text{t}}   \sum_{\mu\nu}^{N_{i\uparrow}} \Braket{\mu_{i\uparrow}\nu_{i\uparrow}|\mu_{i\uparrow}\nu_{i\uparrow}} - \Braket{\mu_{i\uparrow}\nu_{i\uparrow}|\nu_{i\uparrow}\mu_{i\uparrow}} \nonumber\\
        &~+  \frac{1}{2}\sum_{i\neq\el}^{N_\text{t}} \sum_{\mu\nu}^{N_{i\downarrow}} \Braket{\mu_{i\downarrow}\nu_{i\downarrow}|\mu_{i\downarrow}\nu_{i\downarrow}} - \Braket{\mu_{i\downarrow}\nu_{i\downarrow}|\nu_{i\downarrow}\mu_{i\downarrow}} \nonumber\\ 
        &~+ \sum_{i\neq\el}^{N_\text{t}} \sum_{\mu}^{N_{i\uparrow}}\sum_{\nu}^{N_{i\downarrow}}  \Braket{\mu_{i\uparrow}\nu_{i\downarrow}|\mu_{i\uparrow}\nu_{i\downarrow}} 
        + \sum_{j\neq i,\el}^{N_\text{t}} \sum_{\mu}^{N_{i\uparrow}} \sum_{\nu}^{N_{j\downarrow}}
        \Braket{\mu_{i\uparrow} \nu_{j\downarrow}|\mu_{i\uparrow} \nu_{j\downarrow}}
        \nonumber\\
        &~+ \frac{1}{2} \sum_{j\neq i,\el}^{N_\text{t}} \sum_{\mu}^{N_{i\uparrow}} \sum_{\nu}^{N_{j\uparrow}}
        \Braket{\mu_{i\uparrow} \nu_{j\uparrow}|\mu_{i\uparrow} \nu_{j\uparrow}}
        + \frac{1}{2} \sum_{j\neq i,\el}^{N_\text{t}} \sum_{\mu}^{N_{i\downarrow}} \sum_{\nu}^{N_{j\downarrow}}
        \Braket{\mu_{i\downarrow} \nu_{j\downarrow}|\mu_{i\downarrow} \nu_{j\downarrow}}~, 
    \label{eq:hf_ene_res}
\end{align}
\endgroup
and the variation of the Lagrangian functional with respect to an electronic orbital is written as
\begin{equation}
        \frac{\updelta\mathcal{L}}{\updelta\bra{\kappa_{\el}}}  = \frac{\updelta E}{\updelta\bra{\kappa_{\el}}} - 2 \sum_{\nu}^{N_{\el}/2} \epsilon_{\el,\kappa\nu} \Braket{\updelta\kappa_{\el}|\nu_{\el}} = 0 \, .
\end{equation}
After inserting the derivative of the energy expression of Eq.~(\ref{eq:hf_ene_res}), the restricted-unrestricted electronic Fock operator is obtained as
\begin{align}
    f_{\el} =   &~h_{\el}(1)  
        + \sum_{\nu}^{N_{\el}/2} \left(2 \mathcal{J}_{ \el,\nu}(1) -\mathcal{K}_{ \el,\nu}(1)\right) \nonumber\\
        &+\sum_{i\neq \el}^{N_\text{t}} \left( \sum_\nu^{N_{i\uparrow}} \mathcal{J}_{i\uparrow,\nu}(1)
        + \sum_\nu^{N_{i\downarrow}} \mathcal{J}_{i\downarrow,\nu}(1)\right)~,
\end{align}
where the definitions of electronic Coulomb and exchange operators for the restricted case follow trivially from Eqs.\ (\ref{eq:Coulomb}) and (\ref{eq:Exchange}), respectively.
The restricted-unrestricted nuclear Fock operator has the same structure as the unrestricted operator, see Eq.~(\ref{eq:fockop_uhf}), with the addition of the Coulomb interaction term between quantum nuclei and electrons
\begin{align}
    f_{i\uparrow} =   &~h_i(1)  
        + \sum_\nu^{N_{i\uparrow}} \left( \mathcal{J}_{i\uparrow,\nu}(1) -\mathcal{K}_{i\uparrow,\nu}(1)\right) \nonumber\\
        &+ \sum_\nu^{N_{i\downarrow}}  \mathcal{J}_{i\downarrow,\nu}(1)
        + \sum_{j\neq i,\el}^{N_\text{t}} \sum_\nu^{N_{j\uparrow}} \mathcal{J}_{j\uparrow,\nu}(1)
        + \sum_{j\neq i,\el}^{N_\text{t}} \sum_\nu^{N_{j\downarrow}} \mathcal{J}_{j\downarrow,\nu}(1) \nonumber\\
        &+\sum_{\nu}^{N_{\el}/2} 2 \mathcal{J}_{ \el,\nu}(1)~.
\end{align}

\subsubsection{Restricted-Unrestricted Pople--Nesbet-like Equations}

Since in the restricted determinant, the spatial part of spin-up and spin-down molecular orbitals is identical, the electronic density matrix $\bm{D}_{\el}$ follows from the condition $c_{\el\uparrow,\mu p} =c_{\el\downarrow,\mu p}$ as
\begin{equation}
  D_{\el,pq} = 2 \sum_{\mu}^{N_{\el}/2} c_{ \el,\mu p} c_{ \el,\mu q}~,
  \label{eq:DensityMatrixRestricted}
\end{equation}
while the nuclear density matrix for spin $s$ reads
\begin{equation}
  D_{is,pq} = \sum_{\mu}^{N_{is}} c_{is, \mu p} c_{is, \mu q}~,
  \label{eq:DensityMatrixUnrestricted}
\end{equation}
with Eqs.~(\ref{eq:DensityMatrixRestricted}) and (\ref{eq:DensityMatrixUnrestricted}), the electronic Fock matrix can be expressed in the GTO basis as
\begin{equation}
  \begin{aligned}
    F_{\el,pq} = &~h_{\el,pq} + \sum_{rt}^{L_{\el}}  D_{\el,rt} \left( \Braket{p_{\el} r_{\el}|q_{\el} t_{\el}} - \frac{1}{2}  \Braket{p_{\el} r_{\el}|t_{\el} q_{\el}} 
    \right) \\
    &~+  \sum_{j\neq \el}^{N_\text{t}} \sum_{rt}^{L_j} \left(  D_{j \uparrow,rt} 
        + D_{j\downarrow,rt}  \right) \Braket{p_{\el} r_j|q_{\el} t_j} \, , \\
  \end{aligned}
  \label{eq:FockRestricted}
\end{equation}
and the nuclear counterpart for the spin-up component (the extension to the spin-down component follows trivially) as
\begin{equation}
\begin{aligned}
    F_{i\uparrow,pq} = &~h_{i,pq} + \sum_{rt}^{L_i} 
    \Big[ D_{i\uparrow,rt} \big( \Braket{p_i r_i|q_i t_i} -  \Braket{p_i r_i|t_i q_i} \big)
    + D_{i\downarrow,rt} \Braket{p_i r_i|q_i t_i} \Big] \\
    &~+  \sum_{j\neq i,\el}^{N_\text{t}} \sum_{rt}^{L_j} \left(  D_{j\uparrow,rt} 
        + D_{j\downarrow,rt}  \right) \Braket{p_i r_j|q_i t_j} 
    ~+\sum_{rt}^{L_{\el}}  D_{\el,rt} \Braket{p_i r_{\el}|q_i t_{\el}} \, ,
\end{aligned}
\end{equation}
The nuclear-electronic Pople--Nesbet-like equations in matrix notation read
\begin{align}
 \bm{F}_{\el}(\bm{D}_{\el},\{\bm{D}_{js^\prime}\}) \bm{C}_{\el} &= \bm{S}_{\el}  \bm{C}_{\el}  \bm{E}_{\el},\qquad
 \forall ~j\neq\el,\,s^\prime=\uparrow,\downarrow \,,
 \label{eq:PN1} \\
 \bm{F}_{is} (\bm{D}_{\el},\{\bm{D}_{js^\prime}\})\bm{C}_{is} &= \bm{S}_i \bm{C}_{is} \bm{E}_{is} ,\qquad
 \forall ~ij\neq\el,\,ss^\prime=\uparrow,\downarrow \,.
 \label{eq:PN2}
\end{align}
The solution of this system of equations yields sets of molecular orbitals $\mathcal{B}_{i s} = \{ \ket{\mu_{is}} \}$ with dimensions $\mathrm{dim}(\mathcal{B}_{i s}) = L_i$.
As opposed to the case of electrons in the external field of classical nuclear point charges, for two or more particle types one can choose between different schemes to solve Eqs.~(\ref{eq:PN1})--(\ref{eq:PN2}) self-consistently. 
We solve them sequentially, that is, we first update the density matrix of a given type and then employ it to construct the Fock matrix of the next type.
A single iteration is finished, when the density matrices of all types have been updated once.
Furthermore, we employ the DIIS algorithm \cite{Pulay1980_DIIS} to accelerate convergence.

With the nuclear-nuclear interaction matrix, $\bm{G}_{i\uparrow}$, with elements
\begin{equation}
\begin{aligned}
    G_{i,\uparrow, pq} = &~\sum_{rt}^{L_i} 
    \Big[ D_{i\uparrow,rt} \left( \Braket{p_i r_i|q_i t_i} -  \Braket{p_i r_i|t_i q_i} \right) 
    + D_{i\downarrow,rt} \Braket{p_i r_i|q_i t_i} \Big] \\
    &~+  \sum_{j\neq i,\el}^{N_\text{t}} \sum_{rt}^{L_j} \left(  D_{j\uparrow,rt} 
        + D_{j\downarrow,rt}  \right) \Braket{p_i r_j|q_i t_j}~,
\end{aligned}
\end{equation}
and the electron-nuclear interaction matrix, $\bm{I}_{\el}$, 
\begin{equation}
    I_{\el, pq}  = \sum_{j\neq \el}^{N_\text{t}} \sum_{rt}^{L_j} \left(  D_{j\uparrow,rt} 
        + D_{j\downarrow,rt}  \right) \Braket{p_{\el} r_j|q_{\el} t_j} ~,
\end{equation}
the energy can be expressed as
\begin{equation}
\begin{aligned}
 E =&~ \sum_{i\neq \el}^{N_\text{t}} \sum_{pq}^{L_i} \left(D_{i\uparrow,pq} 
 + D_{i\downarrow,pq}  \right) h_{i,pq} 
 + \frac{1}{2} \sum_{i\neq\el}^{N_\text{t}}  \sum_{pq}^{L_i} \left(D_{i\uparrow,pq} G_{i\uparrow,pq}
 + D_{i\downarrow,pq} G_{i\downarrow,pq} \right)
 \nonumber \\
 &+  \frac{1}{2}\sum_{pq}^{L_{\el}}D_{\el,pq} \left( h_{\el,pq} +F_{\el,pq} + I_{\el,pq} \right) 
 \nonumber \\
 = &~\frac{1}{2}\mathrm{Tr} 
 \left[ \bm{D}_{\el} \left( \bm{h}_{\el} + \bm{F}_{\el} + \bm{I}_{\el}\right) \right] ~+
 ~\sum_{i\neq \el}^{N_\text{t}}\sum_{s=\uparrow,\downarrow} \mathrm{Tr} 
 \left[\bm{D}_{is} \left( \bm{h}_i + \frac{1}{2} \bm{G}_{is} \right) \right]~.
\end{aligned}
\end{equation}

\subsection{Second Quantization}
\label{sec:2nd-quant}

The nuclear-electronic Hamiltonian of Eq.~(\ref{eq:PreBO-Hamiltonian}) can be written in the second quantization formalism as
\begin{equation}
 \mathcal{H} = \sum_i^{N_t} \sum_{\mu\nu}^{L_i}  \sum_{s=\uparrow,\downarrow}
  t^{(i)}_{\mu\nu}~ a^\dagger_{is, \mu}a_{i s,\nu}
  + \frac{1}{2} \sum_{ij}^{N_t} \sum_{\mu\kappa}^{L_i}
  \sum_{\nu\lambda}^{L_j} \sum_{ss^\prime=\uparrow,\downarrow}
  V^{(ij)}_{\mu\nu\kappa\lambda} ~ 
  a^\dagger_{is, \mu} a^\dagger_{js^\prime,\nu}   
  a_{j s^\prime, \lambda } a_{i s, \kappa} ~,
  \label{eq:SQPreBO}
\end{equation}
where $t_{\mu\nu}^{(i)}$ and $V^{(ij)}_{\mu\nu\kappa\lambda}$ are the one- and two-body integrals, respectively, calculated over spatial molecular orbitals.
Moreover, we define the creation, $a_{i s, \mu}^\dagger$, and annihilation operators, $a_{i s, \mu}$, acting on orbital $\mu$ of particle type $i$ with spin $s$ that follow the set of anticommutation relations for fermions
\begin{equation}
 \begin{aligned}
   \{a^\dagger_{i s, \mu}, a^\dagger_{i s^\prime, \nu}\} &= 0 \\
   \{a_{i s, \mu}, a_{i s^\prime, \nu} \} &= 0 \\
   \{a^\dagger_{i s, \mu}, a_{i s^\prime, \nu}\} &= \delta_{\mu\nu}\delta_{ s s^\prime} ~,
 \end{aligned}
 \label{eq:CommutationRulesFermions}
\end{equation}
where $\{\cdot , \cdot\}$ is the anticommutator.
Operators belonging to different particle types commute since they act on different subspaces.

We denote the basis functions of the Fock space as occupation number vectors $ \ket{ \sigma_{i, 1}, \cdots, \sigma_{i, L_i} }$, where $\sigma_{i,\mu}$ is the occupation number of the spatial orbital $\varphi_{i,\mu}$.
The Hamiltonian, $\mathcal{H}$, acts on the nuclear-electronic Fock space, which is spanned by the direct product of the occupation number vectors (ONVs) of all particle types ($\ket{\bm{\sigma}}$)
\begin{equation}
  \ket{\bm{\sigma}} = \ket{ \bm{\sigma}_1} \otimes \ket{\bm{\sigma}_2} \otimes
                      \cdots \otimes \ket{ \bm{\sigma}_{N_{\text{t}}}} 
                    = \ket{ \bm{\sigma}_1, \bm{\sigma}_2, \cdots, \bm{\sigma}_{N_{\text{t}}}} \, .
  \label{eq:nuclear-electronic_ONV} 
\end{equation}
We restrict the presentation to the spin-restricted formalism for spin-$\frac{1}{2}$ fermionic particles.
In this case, an orbital can be either empty, occupied with a spin-up or spin-down particle, or doubly occupied, $\ket{ \sigma_{i,\mu}} \in \{ \ket{0}, \ket{\uparrow}, \ket{\downarrow}, \ket{\uparrow\downarrow} \}$.
The full configuration interaction (FCI) wave function reads in terms of ONVs \cite{Thomas1969_protonic1,Muolo2020_Nuclear,Pavosevic2021_MulticomponentUCC}
\begin{equation}
  \ket{\Psi_{\text{FCI}}}  =  
  \sum_{ {\bm{\sigma}}_1 \bm{\sigma}_2 \cdots {\bm{\sigma}}_{N_{\text{t}}}}^{N_\mathrm{FCI}}
  C_{\bm{\sigma}_1 \bm{\sigma}_2 \cdots \bm{\sigma}_{N_{\text{t}}} } 
  \ket{ \bm{\sigma}_1} \otimes \ket{\bm{\sigma}_2} \otimes\cdots\otimes 
  \ket{ \bm{\sigma}_{N_{\text{t}}}}
  = \sum_{\bm{\sigma}} C_{\bm{\sigma}} |\bm{\sigma}\rangle ~,
 \label{eq:GeneralizedCI-SQ}
\end{equation}
with coefficients $C_{ {\bm{\sigma}}_1{\bm{\sigma}}_2\dots{\bm{\sigma}}_{N_{\text{t}}} }$ that form the CI coefficient tensor. The exact CI tensor is obtained by exact diagonalization of the Hamiltonian in the FCI basis.

\subsection{DMRG as a Nuclear-Electronic FCI Solver}
\label{sec:mc-dmrg}

The FCI space spans all possible ONVs with identical numbers of particles.
Its size grows factorially with the dimension of the molecular orbital basis for each particle type.
Exact diagonalization of $\mathcal{H}$ in this basis is computationally feasible only for systems composed of a few particles.
The DMRG algorithm aims at the FCI solution by approximating the ground and excited states as matrix product states (MPSs) where the CI tensor of Eq.~(\ref{eq:GeneralizedCI-SQ}) is represented by its low-rank tensor-train (TT) factorization:
\begin{align}
 \ket{\Psi_{\text{MPS}}} = \sum_{\bm{\sigma}} 
 \underbrace{ \sum_{a_1,a_2\dots,a_{L-1}}^m
 M_{1,a_1}^{\sigma_1} M_{a_1,a_2}^{\sigma_2} \cdots M_{a_{L-1},1}^{\sigma_L} }_{C_{\bm{\sigma}}}
 \ket{ \bm{\sigma} } ~,
 \label{eq:MPS-Parametrization}
\end{align}

The MPS in Eq.~(\ref{eq:MPS-Parametrization}) is defined by $\mathcal{O}(4Lm^2)$ parameters.
By the TT factorization, $C_{\bm{\sigma}}$ is decomposed into a product of $(L-2)$ rank-3 tensors $M_{a_{i-1},a_i}^{\sigma_i}$ and two rank-2 tensors at sites 1 and $L$, respectively.
The maximum value of the auxiliary indices $a_i$, is known as bond dimension $m$ and controls the accuracy of the MPS factorization.
By systematically increasing $m$, Eq.~(\ref{eq:MPS-Parametrization}) can approximate the FCI wave function to arbitrary accuracy \cite{schollwock2005density,Chan2008_Review,Legeza2008_Book,Zgid2009_Review,Marti2010_Review-DMRG,schollwock2011density,chan2011density,Wouters2014_Review,Kurashige2014_Review,szalay2015tensor,Olivares2015_DMRGInPractice,Yanai2015,Baiardi2020_Review}.

Analogously to the MPS, the Hamiltonian operator can be encoded in a tensor-decomposed form as the so-called matrix-product operator (MPO) being
\begin{equation}
  \mathcal{H} = \sum_{\bm{\sigma},\bm{\sigma}^\prime}^{N_\mathrm{FCI}} \sum_{b_1,b_2\dots,b_{L-1}}^{b_{\text{max}}} 
  H_{1,b_1}^{\sigma_1,\sigma_1^\prime} H_{b_1,b_2}^{\sigma_2,\sigma_2^\prime} \cdots H_{b_{L-1},1}^{\sigma_L,\sigma_L^\prime}
  \ket{\bm{\sigma}} \bra{\bm{\sigma}^\prime} ~.
  \label{eq:mpo}
\end{equation}
In contrast to the MPS, the Hamiltonian representation of Eq.~(\ref{eq:mpo}) is exact;
the algorithm for the MPO construction introduced in Ref.~\citenum{Crosswhite2008_MPO-FiniteAutomata} was applied to electronic \cite{keller2015efficient}, as well as vibrational \cite{Baiardi2017_VDMRG} Hamiltonians.
We recently extended this algorithm to Hamiltonians describing systems composed of multiple types of quantum particles \cite{Muolo2020_Nuclear}.
In this case, each DMRG lattice site is associated with a different particle type, which makes it difficult to impose the proper wave function symmetry.
We employ the symmetry-adapted DMRG algorithm introduced in Ref.~\citenum{Vidal2011_DMRG-U1Symm} to enforce the conservation of the particle number for each particle type.
For indistinguishable fermionic particles, we ensure the proper antisymmetry of the wave function with the Jordan--Wigner transformation \cite{Muolo2020_Nuclear,Veis2016_QuantumPreBO,Pavosevic2021_MulticomponentUCC}, as suggested in Ref.~\citenum{Dolfi2014_ALPSProject}.
Note that, unlike the electronic formulation of DMRG, different so-called Jordan--Wigner filling operators must be defined for each particle type.
As shown in our previous work \cite{Muolo2020_Nuclear}, the Jordan--Wigner counterpart of Eq.~(\ref{eq:SQPreBO}) can be straightforwardly encoded as an MPO as described in Ref.~\citenum{keller2015efficient}.
In general, we divide the DMRG lattice into sublattices, one for each particle type, but we note that the algorithm is also capable of intertwining sites of different particles. 

DMRG optimizes the MPS energy based on the variational principle similar to the alternating least-squares algorithm: in the one-site variant of the DMRG algorithm, the tensors $\mathbf{M}^{\sigma_i}$ are optimized sequentially, starting from the first site and traversing the DMRG lattice. 
The optimization of a single tensor is called micro-iteration, while an iteration of the optimization of the whole MPS (once back and forth) is called macro-iteration or sweep. 
From Eqs.~(\ref{eq:MPS-Parametrization}) and (\ref{eq:mpo}), the minimization of the energy of site $i$ leads to the following eigenvalue problem
\begin{equation}
  \sum_{\sigma_i^\prime} \sum_{a^\prime_{i-1},a^\prime_i}^{m} \sum_{b_{i-1},b_i}^{b_{\text{max}}} 
    L_{a_{i-1},a^\prime_{i-1}}^{b_{i-1}} ~ H_{b_{i-1},b_i}^{\sigma_i,\sigma_i^\prime} ~
    R_{a_i,a^\prime_i}^{b_i} ~ M_{a^\prime_{i-1},a_i^\prime}^{\sigma^\prime_i} = 
    E \, M_{a_{i-1},a_i}^{\sigma_i} ~.
  \label{eq:DMRG_Eigenvalue}
\end{equation}
The rank-3 tensors $L_{a_{i-1},a^\prime_i}^{b_{i-1}}$ and $R_{a_i,a^\prime_i}^{b_i}$ collect the partial contraction of the MPS with the MPO for all sites ranging from 1 to ($i$-1) and from ($i$+1) to $L$, respectively. 
We solve the eigenvalue problem from Eq.~(\ref{eq:DMRG_Eigenvalue}) with the Jacobi--Davidson algorithm.\cite{VanDerVorst2000_JacobiDavidson}
To assess the simulation accuracy, we repeat the DMRG optimization with increasing bond dimension, $m$, and monitor the energy convergence.

\section{Orbital Entanglement in Nuclear-Electronic Wave Functions}
\label{sec:mutinf}

In this section, we introduce an entropy-based metric to quantify inter- and intra-species orbital entanglement based on the one- and two-orbital von Neumann entropy.
To obtain the von Neumann entropy of a given set of orbitals, that is, a single orbital or two orbitals, the Hilbert space is partitioned into the system and an environment, where the system corresponds to the Hilbert space of the target orbital space, while the environment is constructed of all the remaining orbitals \cite{Legeza2003_OrderingOptimization}.
With the single- and two-orbital entropies, we can subsequently obtain the mutual information between pairs of orbitals.
This allows us to employ the Fiedler \cite{Reiher2011_Fiedler} orbital ordering on the sublattices of the particle types to enhance the convergence of the DMRG optimization \cite{Legeza2004_Data,Reiher2005_Convergence}.
To obtain the entropies, we calculate the one- and two-orbital reduced density matrices (1o-RDM and 2o-RDM) as described in the following.

\subsection{One-Orbital Reduced Density Matrices}

A pure state $\rho=\ket{\Psi}\bra{\Psi}$ can be partitioned into two entangled subsystems, referred to as the system ($\sigma_{i,\mu}$) and environment ($\bm{\kappa}$), with the wave function of Eq.~(\ref{eq:GeneralizedCI-SQ}) re-written as\cite{Robin2021_NuclPhysEntanglement}
\begin{equation}
\begin{aligned}
  \ket{\Psi_\mathrm{FCI}} 
  &= \sum_{\bm{\kappa},\sigma_{i,\mu}} C_{\bm{\kappa},\sigma_{i,\mu}}\ \ket{\kappa_1,\dots,\sigma_{i,\mu},\dots,\kappa_{L-1}} \\
  &= \sum_{\bm{\kappa},\sigma_{i,\mu}} \eta_{i,\mu}\ C_{\bm{\kappa},\sigma_{i,\mu}}\ \ket{\kappa_1,\dots,\kappa_{L-1},\sigma_{i,\mu}} \\
  &=\sum_{\bm{\kappa},\sigma_{i,\mu}} \eta_{i,\mu}\ C_{\bm{\kappa},\sigma_{i,\mu}}\ \ket{\bm{\kappa}} \otimes \ket{\sigma_{i,\mu}},
\end{aligned}
    \label{eq:bipartite_expansion}
\end{equation}
where $\eta_{i,\mu}$ is the resulting phase factor from permuting the columns of the Slater determinants.
Here, the system is defined as the $\mu$-th orbital of the $i$-th particle on the DMRG lattice and the environment, $\ket{\bm{\kappa}}$, collects all other sites on the lattice.
The von Neumann entropy is defined as
\begin{equation}
    S=-\mathrm{Tr}\left[ \rho \ln \rho \right],
\end{equation}
where $S=0$ for a pure state and $S= \ln L$ for a maximally mixed state and $L$ is the dimension of the lattice.
We can quantify the degree of entanglement of the orbital $\mu$ of particle type $i$ with the environment according to its single-orbital entropy as
\begin{equation}
  s(1)_{i,\mu} = -\mathrm{Tr}_\mathrm{\bm{\kappa}}\left[ \rho_{i,\mu} \ln \rho_{i,\mu} \right],
  \label{eq:s1_1}
\end{equation}
where the trace is calculated over the states of the environment and $\rho_{i,\mu}$ is the 1o-RDM. 

The 1o-RDM can be straightforwardly calculated by substituting Eq.~(\ref{eq:bipartite_expansion}) in the definition $\rho_{i,\mu} =\mathrm{Tr}_\mathrm{\bm{\kappa}} \ket{\Psi}\bra{\Psi}$.
This is, however, computationally very inefficient, since the CI coefficients are explicitly required. 
A different strategy relies on the following projection operator $\mathcal{O}_{i,\mu}$ \cite{White_2006information}
\begin{equation}
  \left(\mathcal{O}_{i,\mu}\right)_{\sigma \sigma^\prime} = \sum_{\bm{\kappa}} \ket{\bm{\kappa}} \otimes \ket{\sigma_{i,\mu}} \bra{\sigma^\prime_{i,\mu}} \otimes \bra{\bm{\kappa}} = \ket{\sigma_{i,\mu}} \bra{\sigma^\prime_{i,\mu}} \otimes I_{\bm{\kappa}}~,
  \label{eq:transitionOp}
\end{equation}
where we could introduce the identity, $I_{\bm{\kappa}}$, due to the completeness of the $\ket{\bm{\kappa}}$ basis.
The matrix elements of the 1o-RDM can be recast as
\begin{equation}
  \left(\rho_{i,\mu}\right)_{\sigma \sigma^\prime} = \Braket{\Psi|\left(\mathcal{O}_{i,\mu}\right)_{\sigma \sigma^\prime}|\Psi}.
  \label{eq:1ordm}
\end{equation}
The one-orbital von Neumann entropy can then be determined after diagonalizing the 1o-RDM as
\begin{equation}
    s(1)_{i,\mu} = - \sum_{\lambda \in \mathrm{spec}(\rho_{i,\mu})} \lambda\ \mathrm{ln}\ \lambda~.
    \label{eq:s1}
\end{equation}
In the following, we drop the indices $i$ and $\mu$ for the sake of readability, whenever they are not explicitly required. 
We derive the 1o-RDM for a system consisting of multiple spin-$\frac{1}{2}$ fermions of different types.

The single-orbital transition operator, $\mathcal{O}_{\sigma \sigma^\prime}$, is defined by 16 matrix elements, given by all combinations of two basis states $\sigma$ and $\sigma^\prime$. We re-label the transition operators, $\mathcal{O}_{\sigma \sigma^\prime}$, by mapping the indices $\sigma\sigma^\prime$ onto a number $n=1,2,\dots,16$, as introduced in our previous publication \cite{Boguslawski2013_orbital}, according to
\begin{equation}
   \begin{array}{c|cccc}
       \mathcal{O}^{(n)} & \ket{0} & \ket{\downarrow} &  \ket{\uparrow} & \ket{\uparrow\downarrow}\\ 
       \hline
      \bra{0}                   & \mathcal{O}^{(1)} & \mathcal{O}^{(2)} & \mathcal{O}^{(3)}  & \mathcal{O}^{(4)}\\
      \bra{\downarrow}          & \mathcal{O}^{(5)} & \mathcal{O}^{(6)} & \mathcal{O}^{(7)}  & \mathcal{O}^{(8)}\\
      \bra{\uparrow}            & \mathcal{O}^{(9)} & \mathcal{O}^{(10)} & \mathcal{O}^{(11)}  & \mathcal{O}^{(12)}\\
      \bra{\uparrow\downarrow}  & \mathcal{O}^{(13)} & \mathcal{O}^{(14)} & \mathcal{O}^{(15)}  & \mathcal{O}^{(16)}\\ 
   \end{array}
\end{equation}
The matrix elements of the operators are given by $ O^{(n)}_{kl} = \delta_{(l+d(k-1)),n}$ for $n=1,\dots,d^2$. 
The states are labeled by $k,l=1,\dots,d$, and $d$ denotes the dimension of the basis (where $d=4$ for spin-$\frac{1}{2}$ particles).
For the 1o-RDM, only transition operators that conserve the particle number will have a non-zero expectation value over an MPS with a well-defined number of particles.
Additionally, an optimized MPS is an eigenfunction of the $S_{z,i}$ operator for all fermionic particle types $i$. 
Therefore, the matrix elements of operators that do not conserve $S_{z,i}$ will be zero as well.

It is convenient to calculate the expectation value, $\bra{\Psi}\mathcal{O}^{(n)}\ket{\Psi}$, of a transition operator starting from its second-quantized form. 
In the case of spin-$\frac{1}{2}$ fermions, the transition operators in second quantization are presented in Ref.~\citenum{Boguslawski2013_orbital}.
We can obtain the spectrum of the 1o-RDM as
\begin{equation}
    \mathrm{spec}(\rho_{i,\mu}) = \{\langle\mathcal{O}_{i,\mu}^{(1)}\rangle,\langle\mathcal{O}_{i,\mu}^{(6)}\rangle,\langle\mathcal{O}_{i,\mu}^{(11)}\rangle,
    \langle\mathcal{O}_{i,\mu}^{(16)}\rangle\} ~.
\end{equation}

\subsection{Two-Orbital Reduced Density Matrices}

The two-orbital reduced density matrix (2o-RDM) is obtained by applying the derivation presented in the previous paragraph to a system of two orbitals, $\sigma_{i,\mu}$ and  $\tau_{j,\nu}$, that may belong to different particle types.
The matrix elements of the 2o-RDM are conveniently expressed in terms of the two-site transition operator:
\begin{equation}
\left(\mathcal{O}_{ij,\mu\nu}\right)_{\sigma\tau,\sigma^\prime\tau^\prime}
     = \sum_{\substack{\bm{\kappa} \\ \sigma_{i,\mu} \sigma_{i,\mu}^\prime \\ \tau_{j,\nu} \tau_{j,\nu}^\prime}} \ket{\bm{\kappa}} \otimes \ket{\sigma_{i,\mu},\tau_{j,\nu}} \bra{\sigma_{i,\mu}^\prime,\tau_{j,\nu}^\prime} \otimes \bra{\bm{\kappa}}
     =\ket{\sigma_{i,\mu}} \bra{\sigma^\prime_{i,\mu}} \otimes \ket{\tau_{j,\nu}} \bra{\tau^\prime_{j,\nu}} \otimes I_{\bm{\kappa}}~ ~,
    \label{eq:2o_transition_op}
\end{equation}
where $\bm{\kappa}$ refers to the environment states.
The matrix elements of the 2o-RDM consequently read

\begin{equation}
 \left(\mathcal{\rho}_{ij,\mu\nu}\right)_{\sigma\tau,\sigma^\prime\tau^\prime}  
 = \Braket{\Psi|\left(\mathcal{O}_{ij,\mu\nu}\right)_{\sigma\tau,\sigma^\prime\tau^\prime}|\Psi}.
    \label{eq:2ordm}
\end{equation}
If the dimension of the basis at site $\mu$ of particle type $i$ is $d$ and at site $\nu$ of particle type $j$ is $d^\prime$, then the size of the basis of $\rho_{ij,\mu\nu}$ will be $dd^\prime$.
Consequently, $\rho_{ij,\mu\nu} \in \mathbb{R}^{d d^\prime \times d d^\prime}$. In the case of two spin-$\frac{1}{2}$ particle types, $dd^\prime =16$. 

The matrix elements of the 2o-RDM can be expressed in terms of products of single-orbital transition operators, $\mathcal{O}_{i,\mu}^{(n)}$ and $\mathcal{O}_{j,\nu}^{(m)}$, of the one-orbital density matrix acting on site $\mu$ and $\nu$, respectively, defined in Eq.~(\ref{eq:transitionOp}). 
The elements of $\rho_{ij,\mu\nu}$ are written as
\begin{equation}
    \left(\rho_{ij,\mu\nu}\right)_{kl,pq} = \mathcal{O}_{i,\mu}^{(n)} \mathcal{O}_{j,\nu}^{(m)}~,
    \label{eq:2rdmMatElem}
\end{equation}
and mediate a transition from state $\ket{p,q}$ to $\ket{k,l}$.
Here, $n=p+d(k-1)$, $m=q+d^\prime(l-1)$, where $p$ and $k$ label states of site $\mu$, and $q$ and $l$ label states of site $\nu$.
As for the 1o-RDM, the 2o-RDM matrix must conserve both the number of particles and the value of the total spin projected onto the $z$-axis, $S_z$.

We first consider the canonical case of the 2o-RDM for sites belonging to the same spin-$\frac{1}{2}$ fermionic particle type $i$. 
Here, the matrix decomposes into nininee different blocks, $\rho_{ii,\mu\nu}^{N,S_z}$, and hence simplifies to
\begin{equation}
    \rho_{ii,\mu\nu} = \rho^{0,0}_{ii,\mu\nu} \oplus \rho^{1,-\frac{1}{2}}_{ii,\mu\nu }\oplus \rho^{1,\frac{1}{2}}_{ii,\mu\nu} \oplus \rho^{2,-1}_{ii,\mu\nu }\oplus \rho^{2,0}_{ii,\mu\nu}  \oplus \rho^{2,1}_{ii,\mu\nu }\oplus \rho^{3,-\frac{1}{2}}_{ii,\mu\nu} \oplus \rho^{3,\frac{1}{2}}_{ii,\mu\nu }\oplus \rho^{4,0}_{ii,\mu\nu}~,
\end{equation}
where $N$ is the particle number.
For the matrix elements of the blocks $\rho_{ii,\mu\nu}^{N,S_z}$, expressed with single-orbital transition operators, $\mathcal{O}^{(n)}_{i,\mu}$, we refer to Ref.~\citenum{Boguslawski2013_orbital}.

Due to particle number conservation, the 2o-RDM for orbitals belonging to particles of different types is diagonal with elements consisting of all combinations of the diagonal transition operators from the respective one-orbital density matrices. 
Therefore, the spectrum of the 2o-RDM for two different fermions of particle types $i$ and $j$ reads
\begin{equation}
    \mathrm{spec} (\rho_{ij,\mu\nu}) = \{ \langle \mathcal{O}_{i,\mu}^{(n)}\mathcal{O}_{j,\nu}^{(m)} \rangle: n,m=1,6,11,16 \}~.
\end{equation}
The two-orbital entropy is calculated as
\begin{equation}
    s(2)_{ij,\mu\nu} = - \sum_{\lambda \in \mathrm{spec}(\rho_{ij,\mu\nu})} \lambda\ \mathrm{ln}\ \lambda ~.
\end{equation}
where $\lambda$ are the eigenvalues of the 2o-RDM matrix, in our case calculated for an optimized MPS.
The mutual information between orbitals $\mu$ and $\nu$ can then be obtained according to
\begin{equation}
    I_{ij,\mu\nu} = \frac{1}{2} \left( s(2)_{ij,\mu\nu} - s(1)_{i,\mu} - s(1)_{j,\nu} \right)~.
\end{equation}

\section{Numerical Results}
\label{sec:results}

In this section, we calculate ground-state energies, vibrational transition frequencies 
and proton densities of HeHHe$^+$ and HCN. 
While the electronic molecular orbitals are approximated with the atomic basis by Pople \cite{Pople1980_self}, the Karlsruhe basis sets \cite{Ahlrichs2005_def2}, and the correlation consistent basis sets \cite{Ahlrichs2005_def2,Dunning1989_gaussian},
the protonic basis (PB) is chosen as PB4-D = \{4s3p2d\}, PB4-F1 = \{4s3p2d1f\}, and PB5-G = \{5s4p3d2f1g\} that were recently introduced by Hammes-Schiffer and coworkers \cite{Hammes-Schiffer2020_NuclearBasis}.
The Hammes-Schiffer PB Gaussian width parameters were optimized to reproduce the ground-state energy and proton density, calculated with CCSD, and the vibrational excitation energies, calculated with time-dependent nuclear-electronic HF, of HCN and FHF$^-$, with reference data calculated with the Fourier grid Hamiltonian\cite{marston1989_fgh,Hammes-Schiffer2000_fgh} (FGH) method.

In the following, we indicate the combined electronic (X) and protonic (Z) basis sets as [$\el$:(X),$\pr$:(Z)].
For cases in which we use different electronic basis sets (X) and (Y) for atoms A and B, respectively, we refer to the corresponding notation as [$\el$:A(X)B(Y),$\pr$:(Z)].
We calculated nuclear-electronic HF energies and orbitals with our implementation of the theory presented so far that exploits the DIIS algorithm\cite{Pulay1980_DIIS} to enhance the convergence of the SCF iterations and relies on the \texttt{Libint} library for integral evaluation \cite{Libint2}.

The NEHF molecular orbital integrals serve to construct the Hamiltonian MPO, as outlined in Sec.~\ref{sec:mc-dmrg}.
Depending on the number of particles and the basis set size, which is equal to the DMRG lattice size $L$, the bond dimension $m$ can be tuned to reach the desired compromise between accuracy and computational effort.
The nuclear-electronic DMRG algorithm is implemented in the open-source \texttt{QCMaquis} program \cite{QCMaquis}.
If not stated otherwise, the DMRG lattice is constructed with NEHF orbitals which are sorted with the canonical ordering, i.e., with increasing NEHF orbital energy, and it is divided into separate sublattices, one for each particle type. All DMRG calculations were performed with the two-site formulation 
to enhance the DMRG convergence\cite{Hubig2015_SSDMRG}.

However, the optimization may converge to local minima of the energy functional with this setup.
In these cases, we improve DMRG convergence with: (i) the Fielder ordering \cite{Legeza2003_OrderingOptimization,Reiher2011_Fiedler}, or (ii) NOs.
The former is a solution to the problem of inefficient optimization when pairs of highly entangled orbitals end up at a distance on the lattice.
In fact, this will be the case when orbitals are sorted with canonical ordering.
The Fielder ordering is based on the mutual information between orbital pairs on the lattice and optimizes the ordering based on the assumption that pairs of entangled orbital should be close to each other on the lattice \cite{Legeza2003_OrderingOptimization}. In this work, we optimize the ordering separately for the protonic and electronic sublattices.
We note that the Fiedler ordering requires an additional initial optimization of the MPS, based on which the mutual information is calculated. 
In our previous work \cite{Stein2016_AutomatedSelection}, we have shown that the mutual information is sufficiently converged within less than eight sweeps and with a moderate bond dimension of $m=250-500$.
Additionally, NEHF-DMRG convergence can be enhanced also by constructing the MPS based on NOs.
Under the assumption that the main features of the 1o-RDM do not change qualitatively with the bond dimension $m$, we obtain the approximate the NOs from a low-$m$ nearly-converged MPS.
This procedure enables us to screen and sort the orbitals based on their natural orbital occupation numbers (NOONs). It is known \cite{lowdin1955quantum,davidson1972_naturalOrbitals} that the CI expansion converges more rapidly with NOs and that the orbitals that have a vanishing occupation number contribute negligibly to the CI expansion. Hence, those NOs can be screened out.  
Moreover, the NOs can significantly reduce the multireference character of nuclear-electronic wave functions, as was observed with the HBCI method \cite{Brorsen2020_SelectedCI-PreBO,Brorsen2020_multicomponentCASSCF}.
We note here that the Fiedler ordering does not improve upon the ordering of NOs based on their NOONs.
For the evaluation of the NOs, see Appendix \ref{sec:appendixDens}.

\subsection{HeHHe$^+$ Molecular Ion}
\label{subsec:HeHHe+_ene}

The helium nuclear point charges are placed at the equilibrium position of the electronic potential energy surface evaluated with all-electron CCSD \cite{Brorsen2020_SelectedCI-PreBO}, that is, at $\pm1.74535105\ \mathrm{bohr}$ on the $x$-axis.
The NEHF-SCF ground-state energies are shown in Tab.~\ref{tab:hehhe_hf_basis} for different combinations of electronic and protonic basis sets. 
We note that by increasing the size of the latter, from PB4-D (25 functions) to PB4-F1 (35 functions) to PB5-G (69 functions), the energy does not decrease monotonically. 
Most likely, this effect is a consequence of the PB basis sets that are optimized not only with respect to the CCSD energy, but also with respect to proton densities and vibrational excitation energies \cite{Hammes-Schiffer2020_NuclearBasis}.

\begin{table}[htbp!]
  \centering 
  \caption{
  NEHF energies for the HeHHe$^+$ molecule with different basis sets for the electrons and the proton.}
  \label{tab:hehhe_hf_basis}
  \begin{tabular}{l c c c c } 
    \hline\hline
      p$^+$/e$^-$          &     def2-SVP      &     def2-TZVP     &      cc-pVTZ   & $\substack{\displaystyle\text{He(cc-pVTZ)}\\ \displaystyle\text{H(cc-pV5Z)}}$   \\ 
    \hline
      PB4-D                &  $-5.771\ 739$   &   $-5.781\ 570$   &   $-5.785\ 569$ & $-5.786\ 434$ \\
      PB4-F1               &  $-5.771\ 735$   &   $-5.781\ 566$   &   $-5.785\ 635$ & $-5.786\ 431$ \\
      PB5-G                &  $-5.771\ 740$   &   $-5.781\ 562$   &   $-5.785\ 638$ & $-5.786\ 434$ \\
  \hline\hline
\end{tabular} 
\end{table}

We investigate the DMRG ground-state energy convergence with respect to the bond dimension $m$ for the [$\el$:(cc-pVTZ),$\pr$:(PB4-D)] basis set. The results shown in Tab.~\ref{tab:hehhe_m} suggest that
for $m\geq500$ the energy is converged to 1~mHa, and with $m\geq1000$ better than $1~\upmu\mathrm{Ha}$.

\begin{table} 
  \centering 
  \caption{NEHF-DMRG energies of HeHHe$^+$ calculated with the [$\el$:(cc-pVTZ)$,\mathrm{p}^+$:(PB4-D)] basis sets, i.e., $L=70$, and with increasing bond dimension $m$.}
  \label{tab:hehhe_m}
  \begin{tabular}{lc}
    \hline\hline 
      $m$       &   Energy/Ha    \\ 
    \hline
      250       &  $-5.881\ 944$ \\
      500       &  $-5.882\ 161$ \\ 
      1000      &  $-5.882\ 226$ \\ 
      1500      &  $-5.882\ 226$ \\ 
    \hline\hline
  \end{tabular}
\end{table}

We report in Tab.~\ref{tab:hehhe_dmrg_basis} the NEHF-DMRG ground and first excited state energies.
For a given electronic basis set, the ground-state energy is converged to 0.1 mHa with the PB4-F1 protonic basis set, and the energy decreases significantly (by 30 mHa) from the def2-SVP to the cc-pVTZ basis set. 
In contrast to the Hartree--Fock energy (Tab.~\ref{tab:hehhe_hf_basis}), we observe that the NEHF-DMRG energy decreases continuously with the proton basis set size, as one would expect for a a class of basis sets designed for correlated calculations.
With the two largest basis sets, i.e., [$\el$:He(cc-pVTZ)H(cc-pV5Z)] for electrons and [$\pr$:(PB4-D)] and [$\pr$:(PB4-F1)] for the proton, we carried out DMRG calculations with NOs by including in the protonic lattice only the orbitals with NOON $>10^{-12}$.
The basis set pruning through NOON accelerates the DMRG convergence by reducing the lattice size $L$.
With the PB4-D basis set, 13/25 orbitals NOs have NOONs below the threshold and for PB4-F1 it was 22/35.
The removal of these NOs allows us to converge in both cases the $m=500$ MPS with $L=112$ and $L=113$, respectively.
We compare our DMRG ground-state energies to data taken from Ref.~\citenum{Brorsen2020_SelectedCI-PreBO} obtained with the HBCI(SDTQ) method \cite{Brorsen2020_SelectedCI-PreBO} which includes for HeHHe$^+$ all excitations.
The lowest energy obtained with HBCI is $-5.887\ 293\ \mathrm{Ha}$, with the [$\el$:He(cc-pVTZ)H(cc-pV5Z),$\pr$:(8s8p8d)] basis set.
For the same electronic basis set, but with the PB4-D basis set for the proton, the converged NEHF-DMRG energy is $-5.887\,271\ \mathrm{Ha}$.
Hence, the resulting energy is 0.022~mHa above the HBCI energy where the latter relied on the same electronic basis but a larger protonic basis, i.e., 80 orbitals against 12 orbitals after pruning (or 25 orbitals before). 
By including $f$-type functions with the PB4-F1 basis set, the final NEHF-DMRG energy is 0.343~mHa below the HBCI energy.
Here, due to pruning, the size of the DMRG lattice was increased only by one site compared to the PB4-D basis set.

\begin{table}
  \centering 
  \caption{NEHF-DMRG ground ($N=0$) and excited ($N=1$) state energies for the HeHHe$^+$ molecule with different basis sets for the electrons and the quantum proton. The bond dimension was $m=1000$. $L_{\el}$ and $L_{\pr}$ are provided in parentheses. The energy obtained with the HBCI method and the [$\el$:He(cc-pVTZ)H(cc-pV5Z),$\pr$:(8s8p8d)] basis set, taken from Ref.~\citenum{Brorsen2020_SelectedCI-PreBO}, is $E=-5.887\ 293\ \mathrm{Ha}$. 
  } 
  \label{tab:hehhe_dmrg_basis}
  \begin{tabular}{l c c c c} 
  \hline\hline
  p$^+$/e$^-$    &     def2-SVP (15)      &     def2-TZVP (18)     &      cc-pVTZ (45)       
                 & $\substack{\displaystyle\text{He(cc-pVTZ)}\\ \displaystyle\text{H(cc-pV5Z)}}$ (100) \\ 
  \hline     
  $E_{N=0}/$Ha   &               &               &               &                 \\ 
  \hline                         
  PB4-D (25)  & $-5.849\,810$ & $-5.861\,894$ & $-5.882\,226$ & $-5.887\,271 ~^{m=500}$ \\ 
  PB4-F1 (35) & $-5.849\,929$ & $-5.862\,012$ & $-5.882\,358$ & $-5.887\,626 ~^{m=500}$ \\ 
  PB5-G (69)  & $-5.849\,945$ & $-5.862\,029$ & $-5.882\,467$ &                         \\
  \hline     
  $E_{N=1}/$Ha   &               &               &               &                 \\ 
  \hline                          
  PB4-D   & $-5.840\,512$ & $-5.852\,474$ & $-5.873\,610$ &             \\
  PB4-F1  & $-5.841\,252$ & $-5.853\,282$ & $-5.875\,386$ &             \\ 
  PB5-G   & $-5.841\,417$ & $-5.853\,493$ & $-5.875\,831$ &             \\ 
  \hline                          
  $\Delta E/$cm$^{-1}$ &         &               &               &             \\
  \hline                          
  PB4-D   &   2041        &    2068       &    1891       &             \\ 
  PB4-F1  &   1904        &    1916       &    1530       &             \\ 
  PB5-G   &   1872        &    1873       &    1452       &             \\ 
  \hline\hline
\end{tabular} 
\end{table}

\subsection{Excited-State Calculations}
\label{subsec:excitedstates}

We calculate the excited states with the state-specific DMRG[ortho] method, which optimizes a given excited state with a conventional DMRG calculation constrained on the space orthogonal to all the lower-energy excited states \cite{keller2015efficient}.
Such a strategy is, in fact, very efficient for optimizing low-energy excited states such as the ones targeted in the present work.
We note, however, that higher-energy excited states could be efficiently targeted by combining the NEHF-DMRG algorithm with other excited-state DMRG solvers; for instance, based on the inverse-power iteration method, on the FEAST \cite{Baiardi2021_Feast} algorithm, or on the shift-and-invert technique \cite{Baiardi2019_HighEnergy-vDMRG}.

The first excited-state energies of HeHHe$^+$ for different protonic and electronic basis sets are shown in Tab.~\ref{tab:hehhe_dmrg_basis}.
By contrast to the ground state, the energy convergence with respect to the protonic basis set size is slower for the excited state, and the PB4-F1 basis set is necessary to converge the energy below 1~mHa.
This is a further indication that the protonic wave function, which is more delocalized for excited states, requires higher angular momentum protonic orbitals or floating functions which are not centered at the BO equilibrium geometry \cite{Muolo2020_Nuclear}. 

\begin{figure}
  \centering
  \includegraphics[width=0.7\textwidth]{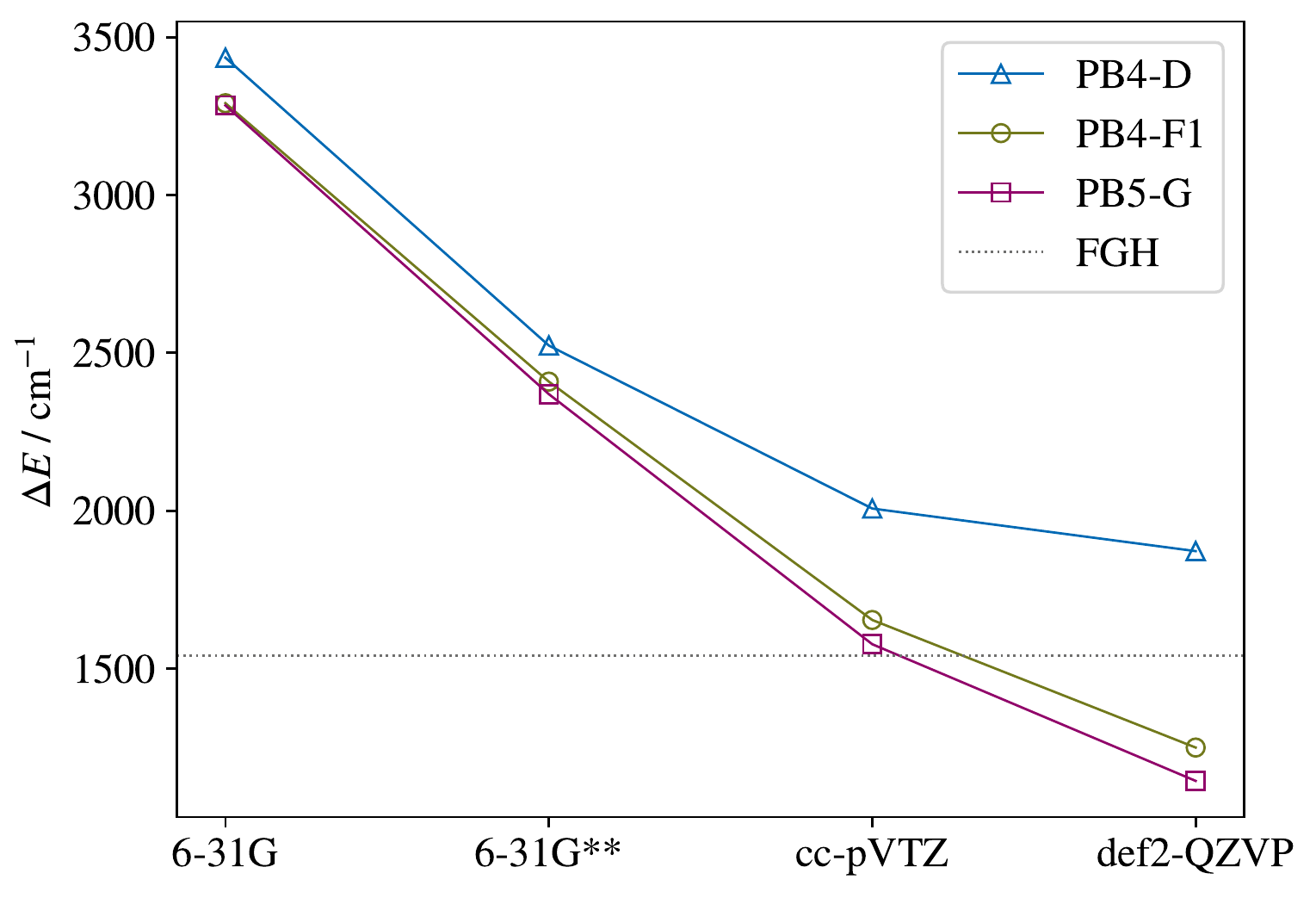}
  \caption{Energy differences (in cm$^{-1}$) of the ground and first excited state of the HeHHe$^+$ molecule
           obtained with the 6-31G electronic basis \cite{Pople1980_self} for He and different electronic and nuclear basis sets for the proton. The He--He distance is $3.40151~\mathrm{bohr}$ and $m=1000$. 
           The reference excitation energy, obtained with FCI for the electrons and the 3D FGH for the proton \cite{marston1989_fgh,Hammes-Schiffer2000_fgh}
           with the 6-31G basis set, is $\Delta_\text{FGH}=1541~\mathrm{cm}^{-1}$ \cite{Hammes-Schiffer2005_NOCI}.}
    \label{fig:hehhe_fgh} 
\end{figure}

We extend the analysis of the excited-state energy convergence to the comparison with the three-dimensional FGH method based on a Born--Oppenheimer PES calculated with FCI based on the 6-31G basis set which contains 2s orbitals at H and He \cite{Pople1980_self}. The results are taken from Ref.~\cite{Hammes-Schiffer2005_NOCI}. 
For the comparison, we employ the 6-31G basis set for the He atoms, which are set apart by $3.40151~\mathrm{bohr}$, and we successively increase the electronic basis set size for the H atom and the PB basis set size. 
The NEHF-DMRG results for bond dimension $m=1000$ are plotted in Fig.~\ref{fig:hehhe_fgh} and listed in Tab.~\ref{tab:hehhe_fgh}.
For the 6-31G electronic basis set, the discrepancy with the FGH energy difference is remarkable: while the latter constructs the delocalized proton wave function based on the 6-31G PES, the NEHF-DMRG wave function does not have diffuse function, which is needed to populate regions of space far from the equilibrium BO geometry, especially for the excited states.
This energy difference vanishes with the Dunning basis set cc-pVTZ which contains up to $d$ orbitals.
For the largest basis set, def2-QZVP, which includes up to $f$ orbitals, NEHF-DMRG converges toward smaller energy differences than FGH.
This result suggests that the error associated with the PES might be greater than a few hundred cm$^{-1}$. 

We highlight here the methodological advantage: while $64^3$ single-point calculations were necessary to approximate the hydrogen PES in the field of fixed Helium atoms on a three-dimensional grid \cite{Hammes-Schiffer2005_NOCI}, a single NEHF-DMRG calculation approximates the proton electron wave function.
Conversely, the downside of nuclear-electronic methods is the need for large basis sets that increases the computational cost of nuclear-electronic post-SCF methods.

\begin{table}
  \centering 
  \caption{Energy differences (in cm$^{-1}$) between ground and first excited state of the HeHHe$^+$ molecule obtained with the 6-31G electronic basis \cite{Pople1980_self} and different PB nuclear basis sets, for a He--He distance of $3.40151~\mathrm{bohr}$ and $m=1000$. 
  The reference excitation energy, obtained with FCI for the electrons and 3D Fourier Grid Hamiltonian (FGH) for the proton \cite{marston1989_fgh,Hammes-Schiffer2000_fgh} with the 6-31G basis set, is $\Delta_\text{FGH}=1541~\mathrm{cm}^{-1}$ \cite{Hammes-Schiffer2005_NOCI}.}
  \label{tab:hehhe_fgh} 
  \begin{tabular}{l c c} 
  \hline\hline
  p$^+$     & $\Delta_\text{DMRG}$ & $\Delta_\text{DMRG} - \Delta_\text{FGH}$  \\
  \hline 
  e$^-$: 6-31G(He)/6-31G(H)     &  &                                           \\ 
  \hline 
  PB4-D     &      $3435$            &                   $1894$                \\
  PB4-F1    &      $3291$            &                   $1750$                \\ 
  PB5-G     &      $3284$            &                   $1743$                \\
  \hline
  e$^-$: 6-31G(He)/6-31G**(H)   &  &                                           \\ 
  \hline 
  PB4-D     &      $2523$            &                    $982$                \\ 
  PB4-F1    &      $2409$            &                    $868$                \\
  PB5-G     &      $2369$            &                    $828$                \\ 
  \hline
  e$^-$: 6-31G(He)/cc-pVTZ(H)   &  &                                           \\ 
  \hline 
  PB4-D     &      $2007$            &                  $466$                  \\ 
  PB4-F1    &      $1654$            &                  $113$                  \\
  PB5-G     &      $1577$            &                  $36$                   \\
  \hline 
  e$^-$: 6-31G(He)/def2-QZVP(H) &  &                                           \\ 
  \hline 
  PB4-D     &      $1872$            &                  $331$                  \\ 
  PB4-F1    &      $1250$            &                  $-291$                 \\
  PB5-G     &      $1145$            &                  $-396$                 \\
  \hline\hline
  \end{tabular}
\end{table}

\subsection{Proton Density and Multireference Effects}
\label{sec:HeHHe+_dens}

\begin{figure}
  \centering
    \includegraphics[width=0.5\textwidth]{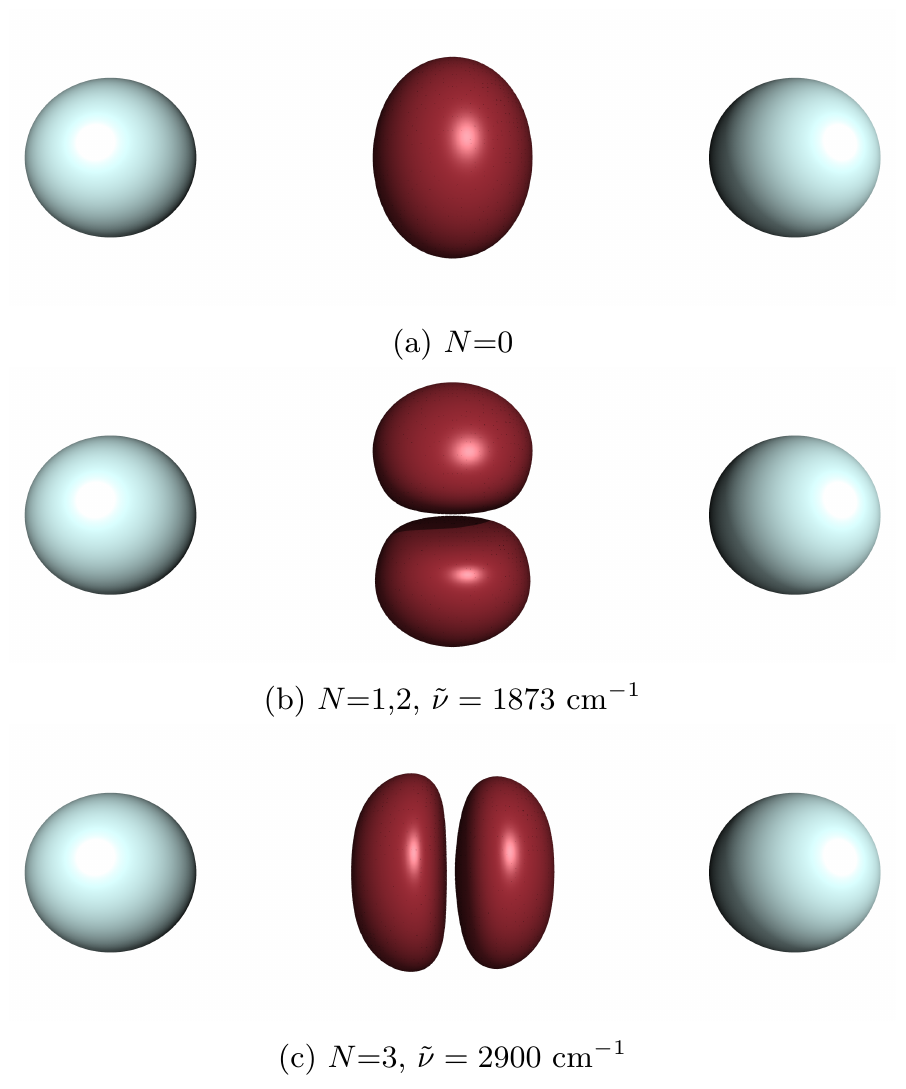}
  \caption{
  Proton densities and vibrational transition frequencies 
  for the (a) ground state, (b) degenerate out-of-plane bends, 
  and (c) antisymmetric stretching mode of the HeHHe$^+$ ion.
  The He--He distance is $3.49\ \mathrm{bohr}$. The results for each state were obtained with
  NEHE-DMRG ($m=1000$), and the [$\el$:(def2-TZVP),$\pr$:(PB5-G)] basis set. 
  The isosurface value chosen is 0.3 a.u.
  } 
  \label{fig:hehehe+_vib}
\end{figure}

\begin{figure}
  \centering
  \includegraphics[width=0.7\textwidth]{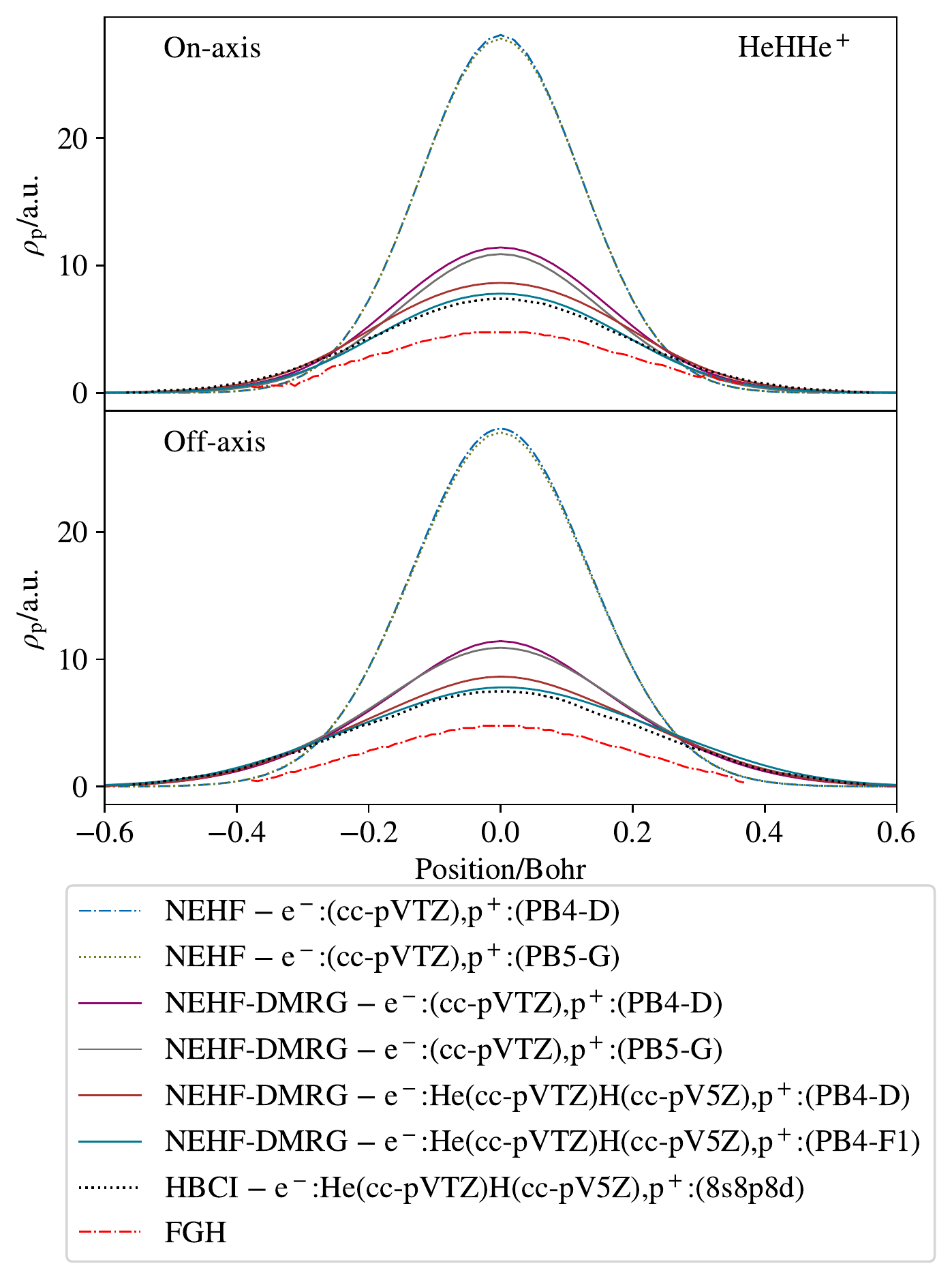}
  \caption{
  NEHF and NEHF-DMRG on and off-axis proton densities for HeHHe$^+$ obtained with the different basis sets  and He nuclei centered at $\pm1.74535105\ \mathrm{bohr}$. 
  The DMRG calculations with [$\el$:(cc-pVTZ),$\pr$:(PB5-G)], [$\el$:He(cc-pVTZ)H(cc-pVQZ),$\pr$:(PB5-G)], [$\el$:He(cc-pVTZ)H(cc-pV5Z),$\pr$:(PB4-D)], and [$\el$:He(cc-pVTZ)H(cc-pV5Z),$\pr$:(PB4-F1)] were carried out with the Fiedler ordering and NOs. $m=1000$ for all calculations but the ones with the cc-pV5Z basis sets, where $m=500$. The HBCI and FGH results are taken from Ref.~\citenum{Brorsen2020_SelectedCI-PreBO}.}
\label{fig:hehhe+_nice_proton_dens}
\end{figure}

In this section, we present one- and three-dimensional proton densities for the lowest vibrational states of HeHHe$^+$ with He--He distance of $3.49\ \mathrm{bohr}$.
For the evaluation of particle densities see Eq.~(\ref{eq:dens2}) in Appendix~\ref{sec:appendixDens}.
The qualitative features of the proton densities for the lowest four states are shown in Fig.~\ref{fig:hehehe+_vib}, and the quantitative analysis of ground state proton density and comparison with reference data are shown in Fig.~\ref{fig:hehhe+_nice_proton_dens}.
For a quantitative comparison of proton densities, we consider on- and off-axis densities (on-axis refers to the bond-axis, on which the proton's basis functions are centered in this context, and off-axis refers to the axis that is perpendicular to the bond axis and passes through the centers of the protonic basis function.
The NEHF proton distributions are much more localized than the NEHF-DMRG ones, with the NEHF density of the proton at the center of mass being $\approx25$ a.u.\ higher than the NEHF-DMRG ones.
Proton overlocalization in single configuration nuclear-electronic wave function is a well known phenomenon that has been investigated for different systems, for example, HeHHe$^+$, HCN, and FHF$^-$, and different basis sets \cite{Hammes-Schiffer2020_Review,Brorsen2020_SelectedCI-PreBO}.
This has usually been taken as an indication that the nuclear-electronic wave function is even qualitatively well described only by a multi-determinantal expansion.

Consequently, a proper inclusion of electron-proton correlation effects through multi-determinantal wave function is crucial to reproduce the delocalized proton density correctly. 
We address the problem of approximating the FCI wave function with the DMRG.
In Fig.~\ref{fig:hehhe+_nice_proton_dens}, the NEHF-DMRG results are compared with HBCI(SDTQ) proton densities obtained with the [$\el$:He(cc-pVTZ)H(cc-pV5Z),$\pr$:(8s8p8d)] basis set and FGH results obtained with the BO-CCSD method (note that the data are obtained from a different geometry and method compared to the FGH data presented above).
Both the HBCI(SDTQ) and FGH results are taken from Ref.~\citenum{Brorsen2020_SelectedCI-PreBO}.
As for the ground-state energy, results show that the protonic basis set size has a relatively small effect on the ground-state proton density, while the He electronic basis set is less decisive for the proton density convergence and properties.
As a consequence, we keep the latter fixed.
For the PB4-D basis set, both the energy and the proton density decrease significantly by increasing the electronic basis at the H nucleus from cc-pVTZ to cc-pV5Z.
For basis sets larger than [$\el$:(cc-pVTZ),$\pr$:(PB4-D)], we carried out DMRG calculations with NOs by including in the protonic lattice only the orbitals with NOON $>10^{-12}$.
Additionally, the best NEHF-DMRG result reproduces the reference FGH proton density qualitatively and matches the HBCI density, which was obtained with the same electronic and a larger protonic basis set.

\begin{figure}
  \centering
    \includegraphics[width=\textwidth]{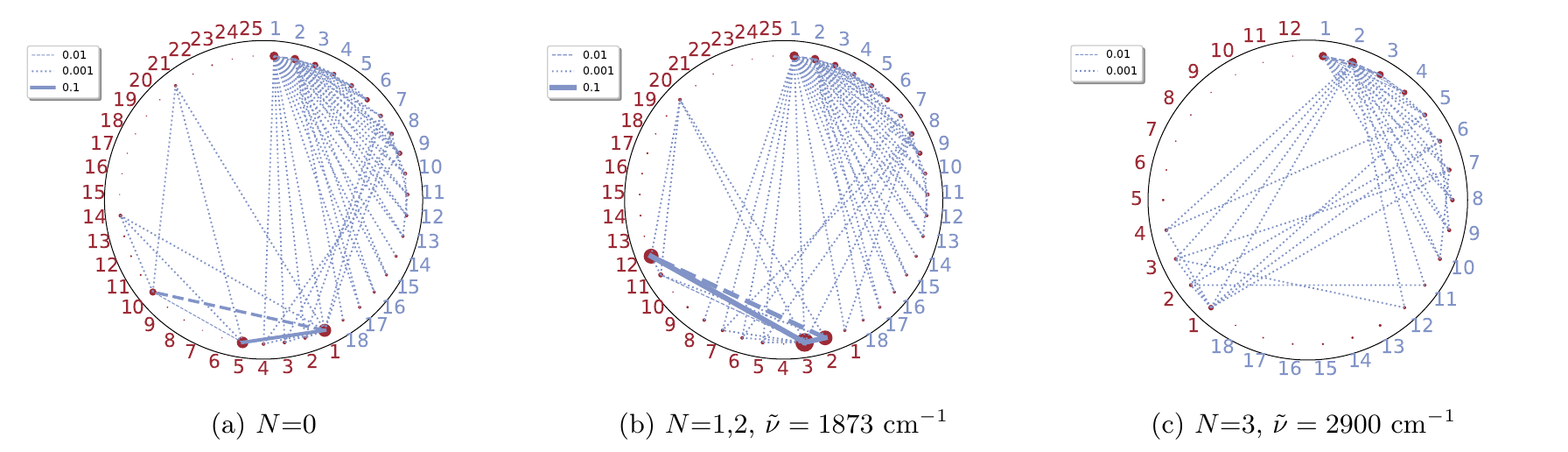}
  \caption{
  Mutual information and single-orbital entropies of the ground state with HF and natural orbitals and the first excited state with HF orbitals of the HeHHe$^+$ molecular ion for a He--He distance of $3.49\ \mathrm{bohr}$, obtained with the basis set [$\el$:(def2-TZVP),$\pr$:(PB4-D)]. Proton orbitals are associated with red numbers, and electron ones with blue numbers. The orbitals are numbered according to increasing orbital energy. }
  \label{fig:hehhe+_mutinf_tz-pb4d}
\end{figure}

We report in Fig.~\ref{fig:hehhe+_mutinf_tz-pb4d} the mutual information between protonic and electronic orbitals for the ground and the first excited state of the HeHHe$^+$ molecule with def2-TZVP and PB4-D basis sets. 
The mutual information is calculated from the MPS optimized with $m$=1000.
We note that the number of strongly entangled electronic orbitals is larger than the number of nuclear ones, even though entanglement is stronger between nuclear orbitals.
Furthermore, the protonic single-orbital entropy decreases significantly with increasing orbital energy, while this is not the case for the electrons.
This explains why the ground-state energy convergence with respect to the protonic basis set size is faster than that of the electronic basis set, as shown in Tab.~\ref{tab:hehhe_dmrg_basis}, and why truncating the protonic basis does not lead to a significant loss in accuracy. 

These observations are complemented by the findings of Brorsen with the HBCI method \cite{Brorsen2020_SelectedCI-PreBO}.
He observed a small $C_0$ coefficient of 0.874 in the full-CI wave function of HeHHe+, built from HF orbitals, which is an indicator for static correlation.
However, by relying instead on natural orbitals, the coefficient is significantly increased to reach 0.986, indicating that static correlation effects are strongly reduced.
Moreover, with NEO-CC methods it was observed \cite{Pavosevic2019_MulticomponentCC} that orbital relaxation effects are of utmost importance, as Pavo{\v{s}}evi{\'c} and Hammes-Schiffer point out in their work on different flavors of the nuclear-electron CC method.
The authors note that, by virtue of Thouless' theorem, the single-excitation operator incorporates orbital relaxation, meaning that some static correlation is taken into account.

\begin{figure}
    \centering
    \includegraphics[width=0.9\textwidth]{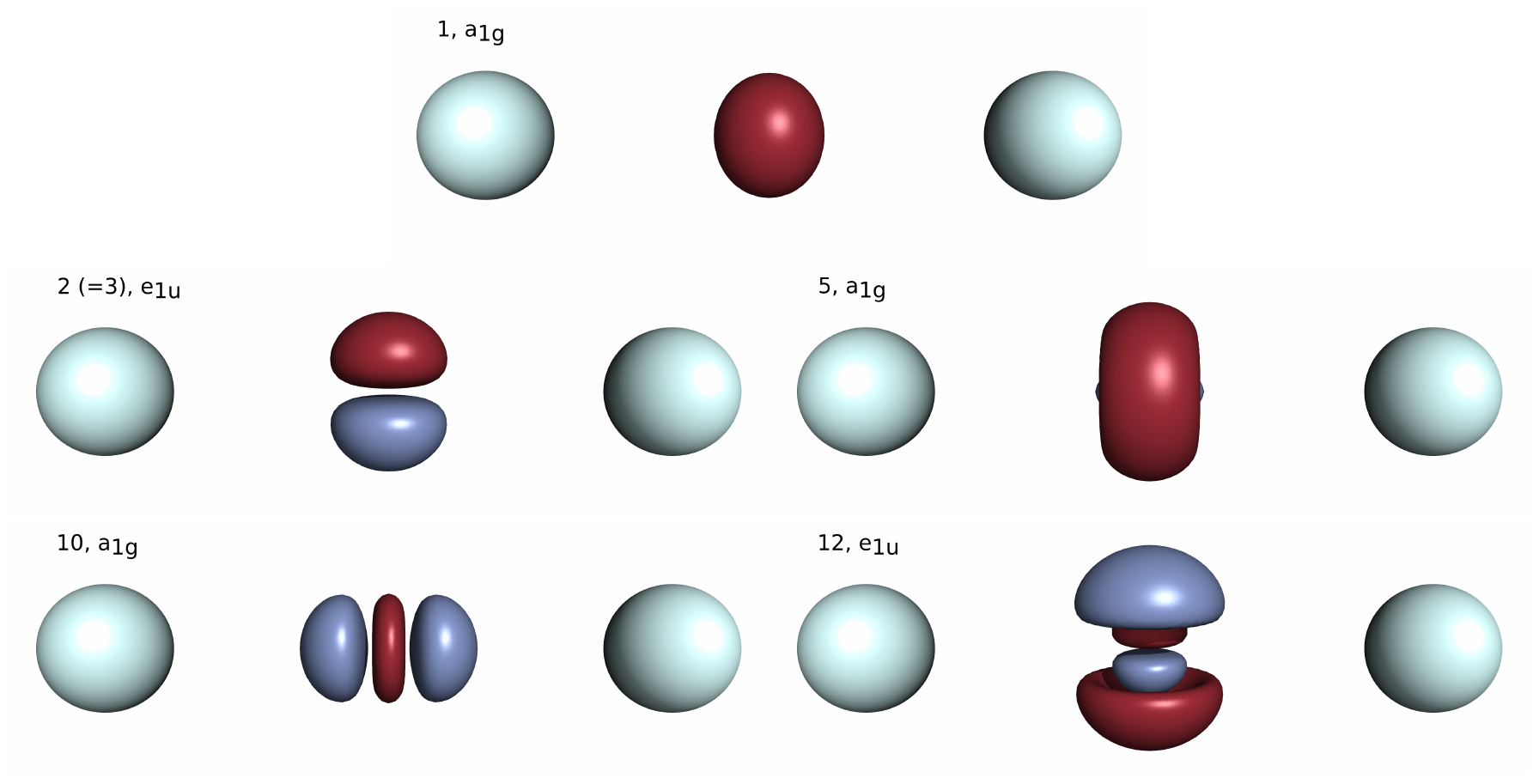}
  \caption{Proton orbitals 1, 2 (=3), 5, 10, and 12 of HeHHe$^+$ for a He--He distance of $3.49\ \mathrm{bohr}$ calculated with the basis set [$\el$:(def2-TZVP),$\pr$:(PB4-D)]. Orbitals are numbered according to increasing orbital energy. The isosurface value chosen is 0.8 a.u.} 
  \label{fig:hehehe+_mos}
\end{figure}

To further investigate proton entanglement effects, Fig.~\ref{fig:hehehe+_mos} depicts the protonic molecular orbitals 1, 2, 3, 5, 10, and 12, of which 1, 5, and 10 are strongly correlated in the ground state.
These orbitals, which are symmetric with respect to the inversion center of the molecule, transform according to the $A_{1g}$ representation of the $D_{\infty h}$ point group.
Hence, they are the only ones that can contribute to the totally symmetric ground-state wave function.
Conversely, the corresponding mutual information diagram becomes less sparse for the first excited state MPS.
The strongly correlated orbitals are in this case 2, 3, and 12, which transform according to the $E_{1u}$ representation of the $D_{\infty h}$ point group, as shown in Fig.~\ref{fig:hehehe+_mos}.

\subsection{The HCN Molecule}
\label{sec:HCN}

\begin{figure}
 \centering
 \includegraphics[width=0.7\textwidth]{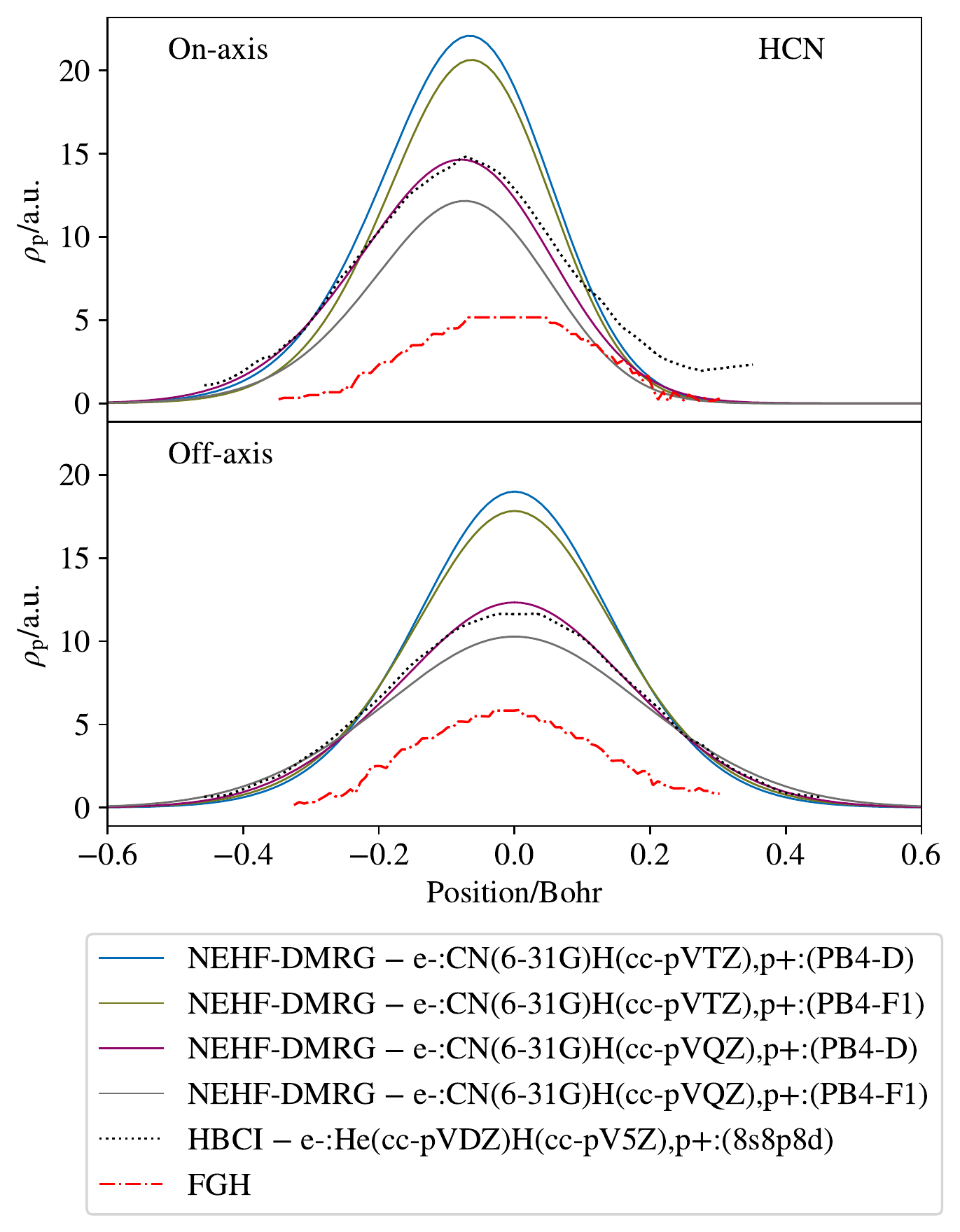}
  \caption{NEHF-DMRG ($m=1000$) on and off-axis proton densities of HCN obtained with the 6-31G electronic basis set for the C and N atoms, the cc-pVTZ and cc-pVQZ electronic bases for the H atom, and the PB4-D and PB4-F1 protonic basis sets.
  The H atom is located at the origin, and the C and N nuclei are located at $1.961157773\ \mathrm{bohr}$ and $4.1649564\ \mathrm{bohr}$ on the x-axis, respectively.
  For comparison, we report the proton density obtained with the HBCI(SDTQ) and the [$\el$:CN(cc-pVDZ)H(cc-pV5Z),$\pr$:(8s8p8d)] basis set, taken from Ref.~\citenum{Brorsen2020_SelectedCI-PreBO}.}
\label{fig:hcn_nice_proton_dens}
\end{figure}

We apply NEHF-DMRG to calculate the ground-state energy and proton density of HCN with electronic and protonic basis sets of increasing size.
We fix the N nucleus at $4.1649564$~bohr, the C nucleus at $1.9611577$~bohr, and the center of the protonic orbitals at the origin.
First, we optimize the MPS, with $m=500$ or $m=750$, and calculate the corresponding mutual information.
We then calculate the NEHF-DMRG energy for larger $m$ values based on the resulting Fiedler ordering. 
We report ground-state energies with various nuclear and electronic basis sets in Tab.~\ref{tab:hcn_energies} and on-axis and off-axis proton densities are shown in Fig.~\ref{fig:hcn_nice_proton_dens}.
We note that the protonic basis set size has a smaller effect on the proton density than the electronic basis set located at the proton's position.
We compare our NEHF-DMRG results with HBCI(SDTQ) energies and proton densities obtained with the basis set [$\el$:CN(cc-pVDZ)H(cc-pV5Z),$\pr$:(8s8p8d)], taken from Ref.~\citenum{Brorsen2020_SelectedCI-PreBO}.
The best NEHF-DMRG density has been obtained with [$\el$:CN(cc-pVDZ)H(cc-pVQZ),$\pr$:(PB4-F1)] and is closer to the FGH reference than HBCI(SDTQ) with a larger basis. 
In fact, NEHF-DMRG, which does not truncate the full-CI wave function, yields a lower absolute energy and a proton density closer to the FGH reference than truncated HBCI(SDTQ) results obtained with a larger basis set.
This suggests that higher-order excitations contribute significantly to the nuclear-electronic wave function, which agrees with the previous observations of Fajen and Brorsen \cite{Brorsen2020_multicomponentCASSCF}
In their work, they observed that dynamic correlation is more important than static correlation for obtaining proton-related properties, which also explains why CC methods are accurate for proton densities.
CCSD proton densities of HCN were published previously \cite{pavovsevic2019_multicomponentCC2}. However, they were obtained with a different geometry and a larger basis set, [$\mathrm{e}^-$:CN(aug-cc-pVTZ)H(aug-cc-pVQZ),$\mathrm{p}^{+}$:(8s8p8d8f)].
A qualitative comparison indicates that the CCSD and DMRG densities are of comparable accuracy with the maxima of both proton densities being at approximately $13\pm 1$~a.u. with the CCSD density being closer to the reference.

\begin{table}[htbp!]
  \centering 
  \caption{
  NEHF-DMRG ($m=1000$) energies for the HCN molecule obtained with different electronic and protonic bases for the H atom.
  The electronic basis set for the C and N atoms is 6-31G.
  $L_{\el}$ and $L_{\pr}$ are provided in parentheses.
  As reference values, the HBCI(SDTQ) energy obtained with [$\el$:CN(cc-pVDZ)H(cc-pV5Z),$\pr$:(8s8p8d)] is $E=-93.17069$~Ha, while the nuclear-electronic CCSD energy is $E=-93.1803$~Ha and the CAS-SCF(10,39)/(1,14) is $E=-93.1857$~Ha, taken from Refs.~\citenum{Hammes-Schiffer2020_NuclearBasis} and \citenum{Brorsen2020_multicomponentCASSCF}, respectively.
  The two latter results are based on the [$\el$:CN(cc-pVDZ)H(cc-pV5Z),$\pr$:(PB4-D)] basis set \cite{Hammes-Schiffer2020_NuclearBasis}.
  We report the electronic and protonic active space sizes in parentheses.} 
  \label{tab:hcn_energies}
  \begin{tabular}{l c c c} 
  \hline\hline
  p$^+$/e$^-$                                            &     
  $\substack{\displaystyle\text{C,N(6-31G)} \\ \displaystyle\text{H(cc-pVTZ)}}$  ($L=33$)  &
  $\substack{\displaystyle\text{C,N(6-31G)} \\ \displaystyle\text{H(cc-pVQZ)}}$  ($L=53$)  &
  $\substack{\displaystyle\text{C,N(cc-pVDZ)} \\ \displaystyle\text{H(cc-pVQZ)}}$ ($L=65$) \\ 
  \hline
  NEHF & & & \\
  \hline
  PB4-D  ($L=25$)  & $-92.792\,643$  & $-92.797\,787$ & $/$   \\
  PB4-F1 ($L=35$)  & $-92.792\,812$ & $-92.798\,135$ & $-92.845\,254$  \\
  \hline
  DMRG & & & \\
  \hline
  PB4-D  ($L=25$)  & $-93.053\,328$ & $-93.069\,328$ & $/$   \\
  PB4-F1 ($L=35$)  & $-93.053\,690$ & $-93.069\,620$ & $-93.196\,494^a$ \\ 
  \hline\hline
\end{tabular} 
\begin{flushleft}
$^a$ $m=1500$.
\end{flushleft}
\end{table}

\begin{figure}
  \centering
\includegraphics[width=\textwidth]{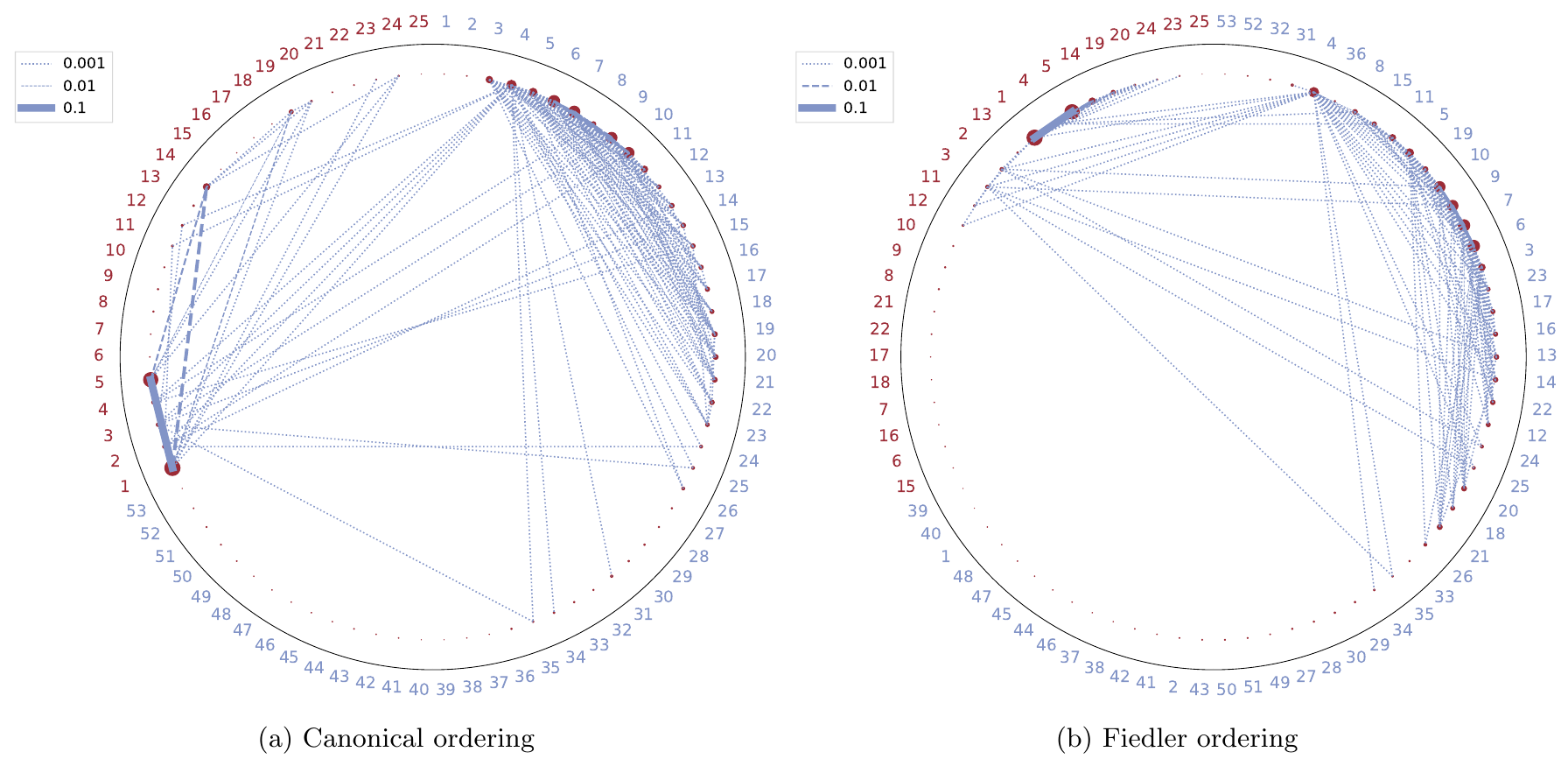}
  \caption{
  Mutual information and single-orbital entropies of the ground state of the HCN molecule obtained with NEHF-DMRG, the [$\el$:CN(6-31G)H(cc-pVQZ),$\pr$:(PB4-D)] basis set, and canonical and Fiedler orbital ordering.
  Protonic orbitals are associated with red numbers, and electronic ones with blue numbers. The orbitals are numbered according to increasing orbital energy.}
  \label{fig:hcn_mutinf}
\end{figure}

\begin{figure}
  \centering
    \centering
    \includegraphics[width=.72\textwidth]{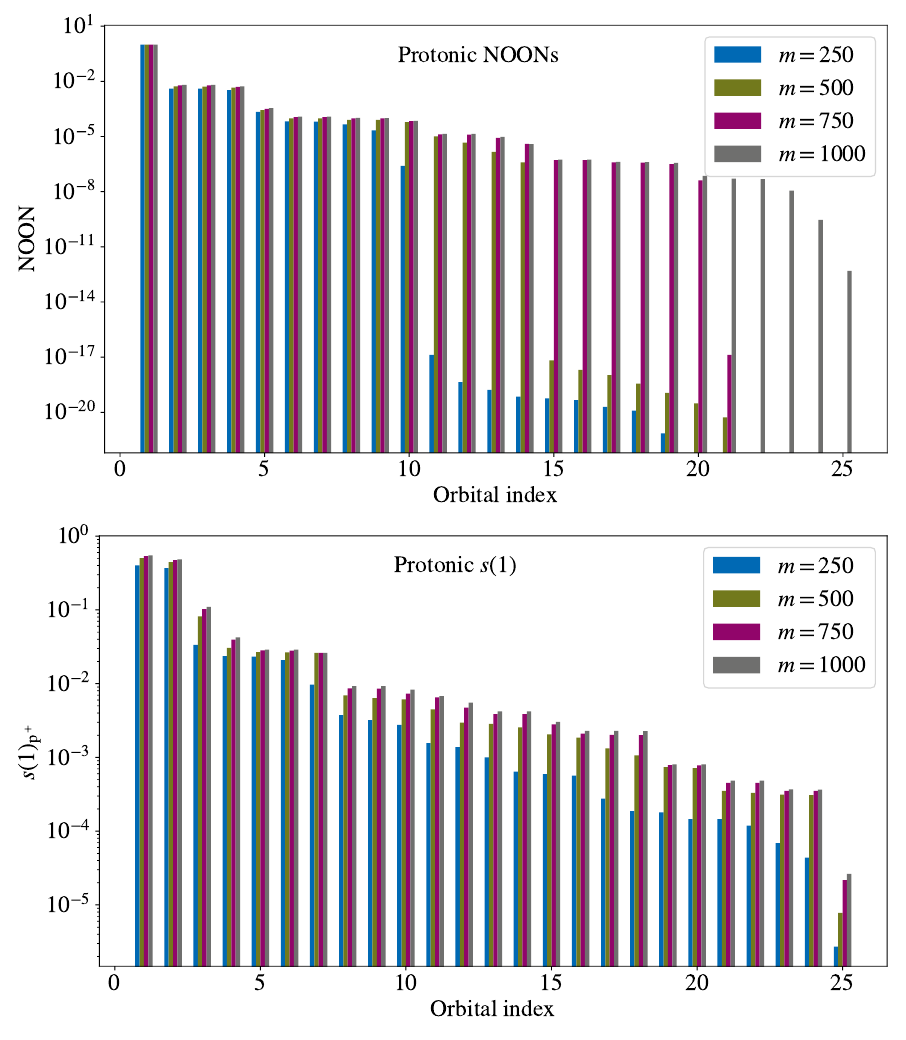}
  \caption{
 Protonic NOONs and single-orbital entropies, $s(1)$, sorted decreasingly, with varying values for the bond dimension $m$ of the orbitals of the HCN molecule, obtained with the basis set [$\el$:CN(6-31G)H(cc-pVQZ),$\pr$:(PB4-D)].
  } 
  \label{fig:hcn_noon_protons}
\end{figure}

In the following, we analyze the nature of correlation effects for HCN based on orbital entanglement measures.
We focus on the relative weight of intra-species (i.e., electron-electron and proton-proton) and inter-species (electron-proton) correlations.
We report in the left panel of Fig.~\ref{fig:hcn_mutinf} the single-orbital entropy and mutual information of the NEHF-DMRG ground state wave function obtained with the [$\el$:CN(6-31G)H(cc-pVQZ),$\pr$:(PB4-D)] basis set, and the canonical orbital ordering.
The orbital-entanglement diagram shows that the inter-species mutual information is consistently lower than 10$^{-3}$, while it is larger than 0.1 for the nuclear orbitals 1 and 5, as well as for the electronic orbitals 6, 7, 8, and 9.
The inter-species entanglement is, in this case, smaller than the intra-species one and this confirms the efficiency of the intra-species Fiedler ordering.
Note that the single-orbital entropy and mutual information decrease monotonically by increasing the electronic orbital index, which is not the case for the nuclear orbitals.
This effect is also observable in the right panel of Fig.~\ref{fig:hcn_mutinf}, where we report the entanglement diagram obtained with the Fiedler ordering.
The orbital ordering optimization leads to a significantly more compact nuclear orbital entanglement, whereas the changes in the electronic one are only minor.
This is confirmed by the relative value of the Fiedler cost function, which decreases from 71.3 to 6.64 for the protons, and only from 121.6 to 89.7 for the electrons.
Fig.~\ref{fig:hcn_mutinf} also shows that many orbitals, especially the nuclear ones, are associated with mutual information smaller than 10$^{-3}$.
The orbital basis could be, therefore, pruned based on the single-orbital entropy, in the spirit of automatic active space electronic-structure algorithms \cite{Stein2016_AutomatedSelection} to reduce the computational cost associated with NEHF-DMRG.
However, this scheme can be efficient only if the orbital entropies can be obtained from a partially converged NEHF-DMRG calculation.
It has been demonstrated\cite{Stein2016_AutomatedSelection} that this is the case for electronic wave functions and, as we show in Fig.~\ref{fig:hcn_noon_protons}, the same holds true for nuclear-electronic wave functions.

In fact, a bond dimension of $m=500$ is sufficient to converge the protonic one-orbital entropy and, therefore, $s(1)$ converge faster with bond dimension $m$ than the energy. 
Conversely, the convergence of the NOON for the protonic orbitals with $m$ is, instead, as slow as for the energy.
These findings suggest that, as already observed for electronic-structure calculations \cite{Stein2016_AutomatedSelection}, s(1) is a more robust correlation metric and therefore it is better suited to automatize nuclear-electronic multireference calculations than NOONs.

\section{Conclusions}
\label{sec:conclusions}

In this work, we harnessed the synergies of a combination of the nuclear-electronic HF method with the DMRG algorithm to accurately represent multi-reference nuclear-electronic wave functions.
We demonstrated that NEHF-DMRG reproduces the reference ground--state total energy and proton density of the HeHHe$^+$ molecular ion.
For HCN, we obtained an accurate approximation to the nuclear-electronic full-CI wave functions efficiently.

We investigated the interplay of nuclear and electronic basis set effects on the energy convergence of NEHF and NEHF-DMRG. 
Our results suggest that the flexibility of the electronic basis set that is located at the position of the protonic basis set is especially crucial to obtain accurate proton-related properties.
Moreover, the size of the protonic basis set becomes significantly more important for excited states, for which the nuclear probability amplitude is more delocalized. 
Finally, the size of the electronic basis sets located at the heavier nuclear point charges strongly affects the total energy of the system.
For both systems, we note that high angular momentum atomic orbitals are necessary to qualitatively reproduce the FGH reference data for proton-related properties.
This is an indication that improved basis sets might help reduce the active orbital space and the multi-reference character of the nuclear-electronic wave functions.

We extended the concepts of orbital entanglement and mutual information to nuclear-electronic wave functions, which provide measures to identify correlated orbitals qualitatively that belong to the same or different particle types. 
We optimized the mapping of the nuclear-electronic orbitals onto the DMRG lattice based on these entanglement measures to improve the DMRG convergence.

The NEHF-DMRG method introduced in the present work can be applied to calculate physical properties other than ground- and excited-state vibrational energies and proton densities.
A challenging example will be proton affinities, which requires the breaking of a donor--H bond and which has already been targeted with multicomponent single-reference methods \cite{Diaz2013_AnyParticleProp,Diaz2013_AnyParticleProp}.

The pre--Born--Oppenheimer time-dependent Schr\"{o}dinger equation can be propagated for a nuclear--electronic wave function ansatz, enabling the description of non-equilibrium non-adiabatic phenomena such as attosecond electron dynamics and hydrogen-transfer reactions.
Recent examples comprise single-reference nuclear-electronic methods such as Hartree--Fock, density functional, and coupled cluster theories\cite{Zhao2020_RealTime,Pavosevic2020_FreqTimeNeoEomCC,HammesSchiffer2020_NeoEhrenfest,Pavosevic2019_MulticomponentCC}.
Analogously, the NEHF-DMRG theory can be straightforwardly extended to time-dependent simulations by combining it with a time-dependent DMRG \cite{Baiardi2019_LargeScale,Baiardi2021_ElectronDynamics} propagation method such as the tangent-space-based approach developed by us for the full quantum chemical Hamiltonian \cite{Muolo2020_Nuclear}.

This work paves the ground for the development of nuclear-electronic DMRG variants that can tackle the correlation problem more efficiently.
In future work, we will develop an algorithm that applies DMRG only to the strongly-correlated orbitals and optimizes the orbitals with nuclear-electronic CAS-SCF \cite{Brorsen2020_multicomponentCASSCF}.
The quantum information metrics introduced here will enable us to fully automate these active-space-based methods, following the strategy of the \texttt{AutoCAS} algorithm \cite{Vera2016_delicate,Stein2016_AutomatedSelection,Stein2017_Chimia,Stein2019_AutoCAS-Implementation} for electronic-structure calculations.
Our future work will also focus on designing the multi-reference extension of nuclear-electronic perturbative\cite{Swalina2004-PreBO_MP2,Pavosevic_OOMP2-Multicomponent} and CC-based approaches,\cite{Pavosevic2019_MulticomponentCC,pavovsevic2019_multicomponentCC2} based on their electronic-structure counterpart,\cite{andersson1990_caspt21,angeli2001_NEVPT2,Yanai2011_DMRGPT2,Roemelt2016_NEVPT2-DMRG,Veis2016_DMRG-TCC,Freitag2017_NEVPT2-DMRG,Morchen2021_TCC} to include efficiently dynamical correlation lacking in active-space approaches.

\section*{Acknowledgments}

R.F. is grateful to the G\"unthard Foundation for a PhD scholarship.
A.M. acknowledges the Swiss National Science Foundation (SNSF) for the funding received through the Early Postdoc Mobility fellowship (grant number P2EZP2\_187994).

\appendix

\section{Particle-Densities and Natural Orbitals}
\label{sec:appendixDens}

The particle density operator for a given particle type $i$ follows from the correspondence principle as
\begin{equation}
  \gamma_i(\mathbf{r}) = \sum_{\mu}^{N_i} \delta(\mathbf{r}-\mathbf{r}_{i,\mu}),
  \label{eq:delta_op}
\end{equation}
where $\delta(\mathbf{r})$ is the Dirac-delta distribution. 
In second quantization, the density operator from Eq.~(\ref{eq:delta_op}) reads
\begin{equation}
  \gamma_i(\mathbf{r}) = \sum_{\mu\nu}^{L_i}\sum_{s=\uparrow,\downarrow} \gamma^{(i)}_{\mu\nu}(\mathbf{r})~a^\dagger_{is, \mu}a_{i,s,\nu},
  \label{eq:dens1}
\end{equation}
where $L_i$ is the number of orbitals of type $i$ and $\gamma^{(i)}_{\mu\nu}(\mathbf{r})$
is the matrix element of the particle density operator between two orbitals, i.e.,
\begin{equation}
    \gamma^{(i)}_{\mu\nu}(\mathbf{r}) = 
    \langle \mu_i |\delta(\mathbf{r}-\mathbf{r}_{i1}) | \nu_i \rangle = \phi_{i,\mu} (\mathbf{r})\phi_{i,\nu} (\mathbf{r}) ~,
    \label{eq:dens_mat_elem}
\end{equation}
where the integration is over $\mathbf{r}_{i1}$.
Considering a nuclear-electronic $N$-body wave function, $\Psi$, the density expectation value is
\begin{equation}
  \langle\gamma_i(\mathbf{r})\rangle = \sum_{\mu\nu}^{L_i} \gamma^{(i)}_{\mu\nu}(\mathbf{r})~\underbrace{\Braket{\Psi| \sum_{s=\uparrow,\downarrow} a^\dagger_{i s, \mu}a_{i s,\nu}|\Psi}}_{\Gamma_{i,\mu\nu}} \, .
  \label{eq:dens2}
\end{equation}
In our approach, the one-particle RDM, $\Gamma_{i,\mu\nu}$, is evaluated by encoding the wave function as an MPS and the operator as an MPO.
We then evaluate the matrix element of the particle density operators from Eq.~(\ref{eq:dens_mat_elem}) in our integral evaluation routine. $\Gamma_{i,\mu\nu}$ is contracted with the matrix elements to give the particle density distribution at a given point $\mathbf{r}$. 
We emphasize here that one should not confuse the one-particle RDM, $\Gamma_{i,\mu\nu}$, with the 1o-RDM and 2o-RDM $\rho_{i,\mu}$ and $\rho_{ij,\mu\nu}$ defined in Eqs.~(\ref{eq:1ordm}) and (\ref{eq:2ordm}), respectively. For the connection between the orbital and particle RDMs see, for example, Ref.~\citenum{boguslawski2015_entanglementInQC}. 

By diagonalizing the one-particle density matrix, one can obtain the natural orbitals which provide a more rapidly convergent CI expansion \cite{lowdin1955quantum,davidson1972_naturalOrbitals}. The diagonalization of the symmetric matrix may be expressed as
\begin{equation}
    \mathbf{U}_i^\mathrm{T}\bm{\Gamma}_i \mathbf{U}_i= \bm{\Lambda}_i,
\end{equation}
where the elements $\lambda_{i,\mu} \in \bm{\Lambda}_i$ are the occupation numbers. 
The natural orbitals are then written as \cite{lowdin1955quantum}
\begin{equation}
 \phi^\mathrm{NO}_{i,\mu} (\mathbf{r}) = \sum_\nu U_{i,\nu\mu} \phi^\mathrm{HF}_{i,\nu}(\mathbf{r}) ,
\end{equation}
where NO and HF refer to natural and Hartree--Fock orbitals, respectively.

\providecommand{\latin}[1]{#1}
\makeatletter
\providecommand{\doi}
  {\begingroup\let\do\@makeother\dospecials
  \catcode`\{=1 \catcode`\}=2 \doi@aux}
\providecommand{\doi@aux}[1]{\endgroup\texttt{#1}}
\makeatother
\providecommand*\mcitethebibliography{\thebibliography}
\csname @ifundefined\endcsname{endmcitethebibliography}
  {\let\endmcitethebibliography\endthebibliography}{}

\end{document}